\RequirePackage{lineno}
\documentclass[aps,prl,twocolumn,showpacs,superscriptaddress,groupedaddress]{revtex4}
\usepackage{graphicx}  
\usepackage{dcolumn}   
\usepackage{bm}        
\usepackage{amssymb}   
\usepackage{amsmath}
\usepackage{appendix}
\usepackage{footnote}
\usepackage{setspace}
\usepackage{enumerate}
\usepackage{lineno}
\usepackage{color}
\usepackage{multirow} 
\usepackage{rotating}
\usepackage{afterpage}
\def \met  {\mbox{$\not\!\!E_T$}}

\begin{document}

\widetext


\title{Constraints on models of the Higgs boson with exotic spin and parity using the full CDF data set\\}

\affiliation{Institute of Physics, Academia Sinica, Taipei, Taiwan 11529, Republic of China}
\affiliation{Argonne National Laboratory, Argonne, Illinois 60439, USA}
\affiliation{University of Athens, 157 71 Athens, Greece}
\affiliation{Institut de Fisica d'Altes Energies, ICREA, Universitat Autonoma de Barcelona, E-08193, Bellaterra (Barcelona), Spain}
\affiliation{Baylor University, Waco, Texas 76798, USA}
\affiliation{Istituto Nazionale di Fisica Nucleare Bologna, \ensuremath{^{jj}}University of Bologna, I-40127 Bologna, Italy}
\affiliation{University of California, Davis, Davis, California 95616, USA}
\affiliation{University of California, Los Angeles, Los Angeles, California 90024, USA}
\affiliation{Instituto de Fisica de Cantabria, CSIC-University of Cantabria, 39005 Santander, Spain}
\affiliation{Carnegie Mellon University, Pittsburgh, Pennsylvania 15213, USA}
\affiliation{Enrico Fermi Institute, University of Chicago, Chicago, Illinois 60637, USA}
\affiliation{Comenius University, 842 48 Bratislava, Slovakia; Institute of Experimental Physics, 040 01 Kosice, Slovakia}
\affiliation{Joint Institute for Nuclear Research, RU-141980 Dubna, Russia}
\affiliation{Duke University, Durham, North Carolina 27708, USA}
\affiliation{Fermi National Accelerator Laboratory, Batavia, Illinois 60510, USA}
\affiliation{University of Florida, Gainesville, Florida 32611, USA}
\affiliation{Laboratori Nazionali di Frascati, Istituto Nazionale di Fisica Nucleare, I-00044 Frascati, Italy}
\affiliation{University of Geneva, CH-1211 Geneva 4, Switzerland}
\affiliation{Glasgow University, Glasgow G12 8QQ, United Kingdom}
\affiliation{Harvard University, Cambridge, Massachusetts 02138, USA}
\affiliation{Division of High Energy Physics, Department of Physics, University of Helsinki, FIN-00014, Helsinki, Finland; Helsinki Institute of Physics, FIN-00014, Helsinki, Finland}
\affiliation{University of Illinois, Urbana, Illinois 61801, USA}
\affiliation{The Johns Hopkins University, Baltimore, Maryland 21218, USA}
\affiliation{Institut f\"{u}r Experimentelle Kernphysik, Karlsruhe Institute of Technology, D-76131 Karlsruhe, Germany}
\affiliation{Center for High Energy Physics: Kyungpook National University, Daegu 702-701, Korea; Seoul National University, Seoul 151-742, Korea; Sungkyunkwan University, Suwon 440-746, Korea; Korea Institute of Science and Technology Information, Daejeon 305-806, Korea; Chonnam National University, Gwangju 500-757, Korea; Chonbuk National University, Jeonju 561-756, Korea; Ewha Womans University, Seoul, 120-750, Korea}
\affiliation{Ernest Orlando Lawrence Berkeley National Laboratory, Berkeley, California 94720, USA}
\affiliation{University of Liverpool, Liverpool L69 7ZE, United Kingdom}
\affiliation{University College London, London WC1E 6BT, United Kingdom}
\affiliation{Centro de Investigaciones Energeticas Medioambientales y Tecnologicas, E-28040 Madrid, Spain}
\affiliation{Massachusetts Institute of Technology, Cambridge, Massachusetts 02139, USA}
\affiliation{University of Michigan, Ann Arbor, Michigan 48109, USA}
\affiliation{Michigan State University, East Lansing, Michigan 48824, USA}
\affiliation{Institution for Theoretical and Experimental Physics, ITEP, Moscow 117259, Russia}
\affiliation{University of New Mexico, Albuquerque, New Mexico 87131, USA}
\affiliation{The Ohio State University, Columbus, Ohio 43210, USA}
\affiliation{Okayama University, Okayama 700-8530, Japan}
\affiliation{Osaka City University, Osaka 558-8585, Japan}
\affiliation{University of Oxford, Oxford OX1 3RH, United Kingdom}
\affiliation{Istituto Nazionale di Fisica Nucleare, Sezione di Padova, \ensuremath{^{kk}}University of Padova, I-35131 Padova, Italy}
\affiliation{University of Pennsylvania, Philadelphia, Pennsylvania 19104, USA}
\affiliation{Istituto Nazionale di Fisica Nucleare Pisa, \ensuremath{^{ll}}University of Pisa, \ensuremath{^{mm}}University of Siena, \ensuremath{^{nn}}Scuola Normale Superiore, I-56127 Pisa, Italy, \ensuremath{^{oo}}INFN Pavia, I-27100 Pavia, Italy, \ensuremath{^{pp}}University of Pavia, I-27100 Pavia, Italy}
\affiliation{University of Pittsburgh, Pittsburgh, Pennsylvania 15260, USA}
\affiliation{Purdue University, West Lafayette, Indiana 47907, USA}
\affiliation{University of Rochester, Rochester, New York 14627, USA}
\affiliation{The Rockefeller University, New York, New York 10065, USA}
\affiliation{Istituto Nazionale di Fisica Nucleare, Sezione di Roma 1, \ensuremath{^{qq}}Sapienza Universit\`{a} di Roma, I-00185 Roma, Italy}
\affiliation{Mitchell Institute for Fundamental Physics and Astronomy, Texas A\&M University, College Station, Texas 77843, USA}
\affiliation{Istituto Nazionale di Fisica Nucleare Trieste, \ensuremath{^{rr}}Gruppo Collegato di Udine, \ensuremath{^{ss}}University of Udine, I-33100 Udine, Italy, \ensuremath{^{tt}}University of Trieste, I-34127 Trieste, Italy}
\affiliation{University of Tsukuba, Tsukuba, Ibaraki 305, Japan}
\affiliation{Tufts University, Medford, Massachusetts 02155, USA}
\affiliation{University of Virginia, Charlottesville, Virginia 22906, USA}
\affiliation{Waseda University, Tokyo 169, Japan}
\affiliation{Wayne State University, Detroit, Michigan 48201, USA}
\affiliation{University of Wisconsin, Madison, Wisconsin 53706, USA}
\affiliation{Yale University, New Haven, Connecticut 06520, USA}

\author{T.~Aaltonen}
\affiliation{Division of High Energy Physics, Department of Physics, University of Helsinki, FIN-00014, Helsinki, Finland; Helsinki Institute of Physics, FIN-00014, Helsinki, Finland}
\author{S.~Amerio\ensuremath{^{kk}}}
\affiliation{Istituto Nazionale di Fisica Nucleare, Sezione di Padova, \ensuremath{^{kk}}University of Padova, I-35131 Padova, Italy}
\author{D.~Amidei}
\affiliation{University of Michigan, Ann Arbor, Michigan 48109, USA}
\author{A.~Anastassov\ensuremath{^{w}}}
\affiliation{Fermi National Accelerator Laboratory, Batavia, Illinois 60510, USA}
\author{A.~Annovi}
\affiliation{Laboratori Nazionali di Frascati, Istituto Nazionale di Fisica Nucleare, I-00044 Frascati, Italy}
\author{J.~Antos}
\affiliation{Comenius University, 842 48 Bratislava, Slovakia; Institute of Experimental Physics, 040 01 Kosice, Slovakia}
\author{G.~Apollinari}
\affiliation{Fermi National Accelerator Laboratory, Batavia, Illinois 60510, USA}
\author{J.A.~Appel}
\affiliation{Fermi National Accelerator Laboratory, Batavia, Illinois 60510, USA}
\author{T.~Arisawa}
\affiliation{Waseda University, Tokyo 169, Japan}
\author{A.~Artikov}
\affiliation{Joint Institute for Nuclear Research, RU-141980 Dubna, Russia}
% ...

\collaboration{CDF Collaboration}
\altaffiliation[With visitors from]{
\ensuremath{^{a}}University of British Columbia, Vancouver, BC V6T 1Z1, Canada,
\ensuremath{^{b}}Istituto Nazionale di Fisica Nucleare, Sezione di Cagliari, 09042 Monserrato (Cagliari), Italy,
\ensuremath{^{c}}University of California Irvine, Irvine, CA 92697, USA,
\ensuremath{^{d}}Institute of Physics, Academy of Sciences of the Czech Republic, 182~21, Czech Republic,
\ensuremath{^{e}}CERN, CH-1211 Geneva, Switzerland,
\ensuremath{^{f}}Cornell University, Ithaca, NY 14853, USA,
\ensuremath{^{g}}University of Cyprus, Nicosia CY-1678, Cyprus,
\ensuremath{^{h}}Office of Science, U.S. Department of Energy, Washington, DC 20585, USA,
\ensuremath{^{i}}University College Dublin, Dublin 4, Ireland,
\ensuremath{^{j}}ETH, 8092 Z\"{u}rich, Switzerland,
\ensuremath{^{k}}University of Fukui, Fukui City, Fukui Prefecture, Japan 910-0017,
\ensuremath{^{l}}Universidad Iberoamericana, Lomas de Santa Fe, M\'{e}xico, C.P. 01219, Distrito Federal,
\ensuremath{^{m}}University of Iowa, Iowa City, IA 52242, USA,
\ensuremath{^{n}}Kinki University, Higashi-Osaka City, Japan 577-8502,
\ensuremath{^{o}}Kansas State University, Manhattan, KS 66506, USA,
\ensuremath{^{p}}Brookhaven National Laboratory, Upton, NY 11973, USA,
\ensuremath{^{q}}Istituto Nazionale di Fisica Nucleare, Sezione di Lecce, Via Arnesano, I-73100 Lecce, Italy,
\ensuremath{^{r}}Queen Mary, University of London, London, E1 4NS, United Kingdom,
\ensuremath{^{s}}University of Melbourne, Victoria 3010, Australia,
\ensuremath{^{t}}Muons, Inc., Batavia, IL 60510, USA,
\ensuremath{^{u}}Nagasaki Institute of Applied Science, Nagasaki 851-0193, Japan,
\ensuremath{^{v}}National Research Nuclear University, Moscow 115409, Russia,
\ensuremath{^{w}}Northwestern University, Evanston, IL 60208, USA,
\ensuremath{^{x}}University of Notre Dame, Notre Dame, IN 46556, USA,
\ensuremath{^{y}}Universidad de Oviedo, E-33007 Oviedo, Spain,
\ensuremath{^{z}}CNRS-IN2P3, Paris, F-75205 France,
\ensuremath{^{aa}}Universidad Tecnica Federico Santa Maria, 110v Valparaiso, Chile,
\ensuremath{^{bb}}The University of Jordan, Amman 11942, Jordan,
\ensuremath{^{cc}}Universite catholique de Louvain, 1348 Louvain-La-Neuve, Belgium,
\ensuremath{^{dd}}University of Z\"{u}rich, 8006 Z\"{u}rich, Switzerland,
\ensuremath{^{ee}}Massachusetts General Hospital, Boston, MA 02114 USA,
\ensuremath{^{ff}}Harvard Medical School, Boston, MA 02114 USA,
\ensuremath{^{gg}}Hampton University, Hampton, VA 23668, USA,
\ensuremath{^{hh}}Los Alamos National Laboratory, Los Alamos, NM 87544, USA,
\ensuremath{^{ii}}Universit\`{a} degli Studi di Napoli Federico I, I-80138 Napoli, Italy
}
\noaffiliation

\date{\today}  

\begin{abstract}

A search for particles with the same mass and couplings as those of the
standard model Higgs boson but different spin and parity quantum numbers
is presented.  We test two specific non-standard Higgs boson hypotheses:
a pseudoscalar Higgs boson with spin-parity $J^P=0^-$ and a graviton-like
Higgs boson with $J^P=2^+$, assuming for both a mass of 125~GeV/$c^2$.  We
search for these exotic states produced in association with a vector boson
and decaying into a bottom-antibottom quark pair. The vector boson is
reconstructed through its decay into an electron or muon pair, or an electron
or muon and a neutrino, or it is inferred from an imbalance in total
transverse momentum.  We use expected kinematic differences between events
containing exotic Higgs bosons and those containing standard model Higgs
bosons.  The data were collected by the CDF experiment at the Tevatron
proton-antiproton collider, operating at a center-of-mass energy of
$\sqrt{s}=1.96$~TeV, and correspond to an integrated luminosity of
9.45~fb$^{-1}$.  We observe no significant deviations from the predictions
of the standard model with a Higgs boson of mass 125~GeV/$c^2$, and
set bounds on the possible rate of production of each exotic state.
\end{abstract}


\maketitle

The observation of a narrow bosonic resonance $H$ with mass near 125~GeV/$c^2$
by the ATLAS~\cite{Aad:2012tfa} and CMS~\cite{Chatrchyan:2012ufa}
Collaborations at the Large Hadron Collider (LHC) in the
$H\rightarrow\gamma\gamma$ and
$H\rightarrow ZZ\rightarrow\ell^+\ell^-\ell^+\ell^-$ decay modes, and the
evidence of such a particle at the Tevatron, primarily in association with
a vector boson and in decays to bottom-antibottom quark
pairs~\cite{Aaltonen:2012qt,Aaltonen:2013kxa}, shifted the focus of
the Higgs boson experimental program to the determination of the properties
of the newly discovered particle.  The central question that needs to be
addressed experimentally is whether only one Higgs boson is sufficient to
explain the observed data.  Specifically, the spin and parity of the Higgs
boson should be established in order to determine if it plays the role
predicted for it by the standard model (SM) of particle physics or if it
represents the first hint of more exotic interactions.

The properties of the new particle observed at the LHC are consistent with
those predicted by the SM for the Higgs boson. The products of cross sections
and branching ratios are as
predicted~\cite{Aad:2013wqa,Aad:2012tfa,Chatrchyan:2013lba,pdghiggs}.
The decays of the new particle to $ZZ^{(*)}$, $\gamma\gamma$, and $WW^{(*)}$
final states, where the asterisk indicates an off-mass-shell {\it Z} or
{\it W} vector boson, provide excellent samples
for testing its spin and parity quantum numbers $J$ and $P$, due to the
measurable angular distributions of the decay
products~\cite{Aad:2013xqa,Chatrchyan:2012jja}, which depend on the quantum
numbers of the decaying particle.  The tests at the LHC
in the bosonic decay channels exclude exotic states with spin and/or parity
different from the SM prediction of $J^P=0^+$ with high confidence level.

At the Tevatron, the primary sensitivity to the Higgs boson comes from modes
in which it is produced via its coupling to vector bosons but decays to a pair
of fermions.  While ATLAS and CMS have reported strong evidence for fermionic
decays of the Higgs boson~\cite{Aad:2015,Chatrchyan:2014vua}, spin and parity
quantum numbers have not been tested in these decays.  As the D0 Collaboration
has shown~\cite{Abazov:2014}, testing the spin and parity of the Higgs boson
at the Tevatron provides independent information on the properties of this
particle.

The Tevatron data can test alternative $J^P$ hypotheses in the {\it WH},
{\it ZH} production modes with $H\rightarrow b{\bar{b}}$, by examining the
kinematic distributions of the observable decay products of the vector boson
and the Higgs-like boson~\cite{Ellis:2012xd}.  Testing the spin and parity
of the Higgs boson in $H\rightarrow b{\bar{b}}$ decays provides independent
information on the properties of this particle.  The models tested are
described in Ref.~\cite{Miller:2001bi}.  For the SM case, Higgs boson
associated production is an {\it S}-wave process (i.e. the {\it VH} system
is in a state with relative orbital angular momentum $L=0$, where {\it V} =
{\it W} or {\it Z}), with a cross section that rises proportionally to the
boson speed $\beta$ close to threshold.  Here $\beta = 2p/\sqrt{s}$, where
$p$ is the momentum of the Higgs boson in the {\it VH} reference frame
and $\sqrt{s}$ is the total energy of the {\it VH} system in its rest
frame~\cite{Miller:2001bi}.  In the $0^-$ case, the production is a $P$-wave
process and the cross section rises proportionally to $\beta^3$.  There are
several possible $J^P=2^+$ models, but for graviton-like
models~\cite{Ellis:2012xd}, the production is in a $D$-wave process, with a
cross section that rises proportional to $\beta^5$.  This dependence of the
cross section on the spin-parity quantum numbers provides good kinematic
leverage for discriminating exotic from SM Higgs boson production, since the
exotic production rate is enhanced faster than the SM one at larger $\beta$,
corresponding to a larger invariant mass of the final state system and higher
momenta of the decay products.  The models studied predict neither the
production cross sections for $p{\bar{p}}\rightarrow WH$, {\it ZH} nor the
decay branching fraction ${\cal B}(H\rightarrow b{\bar{b}})$.  Instead,
the authors suggest~\cite{Ellis:2012xd} to purify a sample of Higgs boson
candidate events and to study the invariant masses of the $Wb{\bar{b}}$ and
$Zb{\bar{b}}$ systems, which differ strongly among the $0^+$, $0^-$, and
$2^+$ models.

The study of the properties of a purified signal sample with minimal
sculpting of the kinematic distributions is effective at the LHC in the
$H\rightarrow ZZ\rightarrow\ell^+\ell^-\ell^+\ell^-$ mode, which has a
signal-to-background ratio $s/b$ exceeding 2:1.  However, this is not the
case for the Tevatron, where the SM Higgs boson searches typically have a
$s/b$ of 1:50~\cite{footnote1}.  With the use of multivariate analyses
(MVAs), small subsets of the data sample can be purified to achieve a $s/b$
ratio of $\approx$ 1:1.  Since the events in these subsets are selected
with MVA discriminants that are functions of the kinematic properties of
signal and background, their distributions are highly sculpted to resemble
those predicted by the SM Higgs boson, and thus are not optimal in testing
alternative models.

The strategy chosen for this Letter is to generalize the CDF searches
for the SM Higgs boson in the
$WH\rightarrow\ell\nu b{\bar{b}}$ mode~\cite{Aaltonen:2012ic}, the
$ZH\rightarrow\ell^+\ell^-b{\bar{b}}$ mode~\cite{Aaltonen:2012id}, and the
$WH+ZH\rightarrow\met b{\bar{b}}$~\cite{footnote2} mode~\cite{Aaltonen:2013js},
where the {\it Z} boson decays into a neutrino pair or the charged lepton from
the {\it W}-boson decay escapes detection.  In the last case,
$ZH\rightarrow\ell^+\ell^- b{\bar{b}}$ events may be reconstructed as
$\met b{\bar{b}}$ events if both leptons fail to meet the identification
criteria.  The generalization involves searches for pseudoscalar ($J^P=0^-$)
and graviton-like ($J^P=2^+$) bosons (denoted $X$ here), using MVA techniques
similar to those developed for the SM searches.  Admixtures of SM and exotic
Higgs particles with indistinguishable mass are also considered, where exotic
and SM production do not interfere due to different spin-parity quantum
numbers.  We set limits on the production rate times the decay branching
ratio ${\cal{B}}(X\rightarrow b{\bar{b}}$ of the exotic boson assuming a
production cross section and decay branching ratio of the exotic boson as
predicted by the SM for the Higgs boson.  We also test the hypotheses of
the exotic models by comparing the data with the predictions.

The CDF~II detector is described in detail
elsewhere~\cite{secvtx,cdfdetector}.  Silicon-strip tracking
detectors~\cite{cdfsilicon} surround the interaction region and
provide precise measurements of charged-particle trajectories in the
range $|\eta|<2$~\cite{coord}.  A cylindrical drift chamber provides
full coverage over the range $|\eta|<1$.  The tracking detectors are
located within a 1.4~T superconducting solenoidal magnet with field
oriented along the beam direction.  The energies of individual particles
and particle jets are measured in segmented electromagnetic and hadronic
calorimeters arranged in a projective-tower geometry surrounding the
solenoid.  Tracking drift chambers and scintillation counters are located
outside of the calorimeters to help identify muon candidates~\cite{cdfmuons}.
The Tevatron collider luminosity is measured with multicell gas Cherenkov
detectors~\cite{CLC}.  The data set used in the analyses reported in this
Letter corresponds to an integrated luminosity of 9.45~fb$^{-1}$.  The data
are collected using a three-level online event selection system (trigger).
The first level, relying on special-purpose hardware~\cite{XFT}, and the
second level, using a mixture of dedicated hardware and fast software
algorithms, reduce the event accept-rate to a level readable by the data
acquisition system.  The accepted events are processed online at the third
trigger level with fast reconstruction algorithms, and recorded for offline
analysis~\cite{cdfl3}.

To predict the kinematic distributions of SM Higgs boson events, we use the
\textsc{pythia}~\cite{pythia} Monte Carlo (MC) program, with CTEQ5L~\cite{cteq}
parton distribution functions (PDFs) of leading order (LO) in the strong
coupling parameter $\alpha_s$.  We scale these MC predictions to the
highest-order cross section calculations available.  To predict the exotic
signal kinematic distributions, we use a modified version of
\textsc{madevent}~\cite{madgraph} provided by the authors of
Ref.~\cite{Ellis:2012xd}.

The predictions for the SM {\it WH} and {\it ZH} cross
sections~\cite{djouadibaglio} are based on the next-to-leading order (NLO)
calculation of {\sc v2hv}~\cite{v2hv} and include next-to-next-to-leading
order (NNLO) quantum chromodynamical (QCD) contributions~\cite{vhnnloqcd},
as well as one-loop electroweak corrections~\cite{vhewcorr}.  In the
predictions for the decay branching fractions of the SM Higgs
boson~\cite{lhcxs,lhcdifferential}, the partial decay widths for all decays
except to pairs of {\it W} and {\it Z} bosons are computed with
\textsc{hdecay}~\cite{hdecay}, and the {\it WW} and {\it ZZ} decay widths
are computed with {\sc prophecy4f}~\cite{prophecy4f}.  The relevant rates
are $\sigma_{WH}=(129.5\pm 9.8)$~fb, $\sigma_{ZH}=(78.5\pm 5.9)$~fb, and
${\cal B}(H\rightarrow b{\bar{b}})=(57.8\pm 1.0)$\%.  The uncertainties
on the predicted branching ratio from uncertainties in the bottom-quark
mass, $\alpha_s$, and missing higher-order effects are estimated in
Refs.~\cite{dblittlelhc,denner11}.

We model SM processes and instrumental backgrounds using data-driven and
MC methods.  Simulated diboson ({\it WW}, {\it WZ}, {\it ZZ}) MC samples
are normalized using the NLO calculations from \textsc{mcfm}~\cite{mcfm}. 
For $t{\bar{t}}$ we use a production cross section of
$7.04\pm 0.7$~pb~\cite{mochuwer}, which is based on a top-quark mass of
173~GeV/$c^2$ and MSTW 2008 NNLO PDFs~\cite{MSTW}.  The single-top-quark
production cross section is taken to be $3.15\pm 0.31$~pb~\cite{kidonakis_st}.
The normalization of the {\it Z}+jets and {\it W}+jets MC samples is taken
from \textsc{alpgen}~\cite{Mangano:2002ea} corrected for NLO effects, except
in the case of the $WH\rightarrow\ell\nu b{\bar{b}}$ search.  The normalization
of the {\it W}+jets MC sample in the $WH\rightarrow\ell\nu b{\bar{b}}$ search,
and the normalization of the instrumental and QCD multijet samples in all
searches, are constrained from data samples where the expected $s/b$ ratio
is several orders of magnitude smaller than in the search samples.  The
quality of background modeling is shown in final-state invariant mass
distribution plots included in the Supplemental Material to this Letter,
which show good agreement with the data in all cases.

The analyses used to search for the exotic pseudoscalar and graviton-like
Higgs bosons are modifications of the searches for the SM Higgs boson,
optimized for separating the exotic signals from both the SM background
sources and the possible SM Higgs boson signal.  They use the most recent
and efficient CDF algorithm, HOBIT~\cite{Freeman:2012uf}, for identifying
jets from the hadronization of bottom quarks ($b$-tagging).  HOBIT is a
multivariate classifier that uses kinematic properties of reconstructed
trajectories of charged particles (tracks) associated with displaced vertices,
the impact parameters of the tracks, and other characteristics of reconstructed
groups of collimated particles (jets) that help separate $b$-jets from
light-flavored jets.  The HOBIT classifier does not perform well for jets
with $E_{\rm{T}}>200$~GeV and the data-based calibration procedures associated
with it suffer from greater uncertainties in this kinematic region.  We
therefore do not tag jets with $E_{\rm{T}}>200$~GeV.  The same tight (T)
and loose (L) tag requirements are used as in the SM Higgs analyses.

In each final state, the search channels are subdivided according to the
number of jets, the lepton category, and the $b$-tag category.  The
$WX\rightarrow\ell\nu b{\bar{b}}$ events are divided into 15 subchannels,
corresponding to the TT, TL, 1T, LL, and 1L tagging categories of the two
jets, for each lepton category: central leptons (electrons or muons),
forward electrons, and isolated-track leptons.  The
$ZX\rightarrow\ell^+\ell^- b{\bar{b}}$ events are divided into 16 subchannels,
corresponding to the TT, TL, 1T, LL tagging categories in the two- and
three-jet final states, separately for
$Z\rightarrow e^+e^-$ and $Z\rightarrow\mu^+\mu^-$ events.
The $WX+ZX\rightarrow\met b{\bar{b}}$ events are divided into 6 subchannels
corresponding to the  TT, TL, and 1T tagging categories in 2-jet and 3-jet
final states.  A total of 37 analysis channels are defined.  The expected
and observed event yields in all channels are summarized in
Table~\ref{tab:yields}, summed over lepton, jet, and $b$-tag categories.

Two discriminant functions are defined for each subchannel, one to separate
the exotic Higgs boson signal (separately defined for the $0^-$ and the $2^+$
signals) from the backgrounds, and the other as the discriminant used in the
search for the SM Higgs boson.  For the $ZX\rightarrow\ell^+\ell^- b{\bar{b}}$
analysis, only the exotic discriminant is used.  The exotic signal
discriminants have either $M_{Vb{\bar{b}}}$ (the invariant mass of the
final-state system) among their input variables or $H_T$ (the sum of all
transverse energies reconstructed in the final state, including muon energies
and $\met$).  Distributions of the discriminant functions for all search
channels are shown for the data and simulation in the Supplemental Material
to this Letter.  Since the events are primarily classified to test for the
exotic models, the SM Higgs interpretation of the data will not be the same
as in the searches optimized for the SM Higgs boson.

\begin{table}[htb]
\begin{center}
\caption[]{Expected and observed event yields for all channels.  The difference
between the $0^-$ and $2^+$ exotic yields is due to different signal
acceptances.}
\label{tab:yields}
\begin{tabular}{lccc}\hline\hline
Process        & $\ell^+\ell^- b{\bar{b}}$ & $\ell\nu b{\bar{b}}$ & $\met b{\bar{b}}$ \\\hline
$V+X_{0^-}$     & 8$\pm$1                   & 49$\pm$4             & 81$\pm$6                 \\\hline
$V+X_{2^+}$     & 7$\pm$1                   & 43$\pm$4             & 65$\pm$5                 \\
$VH$           & 7$\pm$1                   & 33$\pm$3             & 40$\pm$3                 \\\hline
$V$+jets       & 820$\pm$141               & 23323$\pm$2860       & 9193$\pm$2273            \\
Dibosons       & 72$\pm$11                 & 1288$\pm$148         & 544$\pm$66               \\
Top            & 222$\pm$22                & 2053$\pm$211         & 1935$\pm$164             \\
QCD            & 58$\pm$21                 & 2406$\pm$603         & 16283$\pm$1447           \\
Total bkg      & 1172$\pm$272              & 29070$\pm$3037       & 27956$\pm$3188           \\\hline
Observed       & 1182                      & 26337                & 28518                    \\\hline\hline
\end{tabular}
\end{center}
\end{table}

\begin{figure}[htb]
 \begin{centering}
 \includegraphics[width=0.67\linewidth]{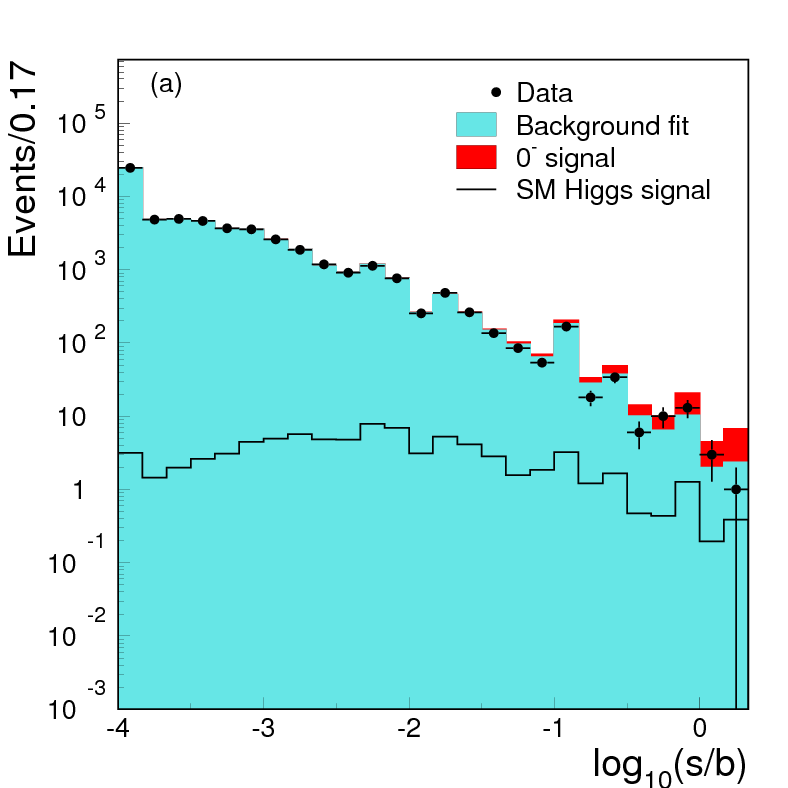}
 \includegraphics[width=0.67\linewidth]{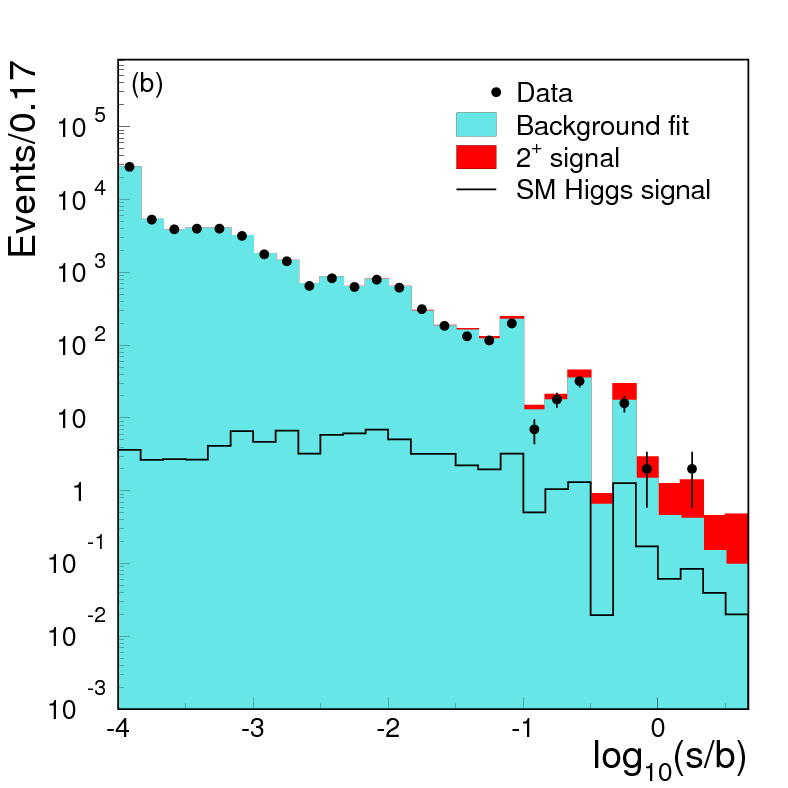}
 \caption{\label{fig:sb}
Distributions of $\log_{10}(s/b)$ for the data from all contributing
Higgs boson search channels for a boson mass of 125~GeV/$c^2$ for (a) the
$0^-$ search and (b) the $2^+$ search.  The observed numbers of events are
represented by the points, and the expected exotic signals are shown as
histograms with $\mu_{\rm{exotic}}=1$ stacked on top of the backgrounds, which
are fit to the data within their systematic uncertainties.  The expected $s/b$
ratios of the exotic signal over the background yield are used to rank analysis
bins.  The background predictions do not include the contributions from the
SM Higgs boson, which are shown as separate histograms, not stacked.  The
error bars shown on the data correspond to the square root of the observed
data count.  Underflows and overflows are collected into the leftmost and
rightmost bins, respectively.}
 \end{centering}
 \end{figure}

To summarize the data in the large number of contributing channels, we follow
Ref.~\cite{Aaltonen:2013kxa}.  We sum the contents of bins with similar $s/b$
ratios over the output histograms of all channels.  Fig.~\ref{fig:sb} shows
the comparison of the data with the best-fit background predictions and the
summed signals, separately for the SM Higgs and exotic boson signals.  The
signal strength modifier is denoted by $\mu_{\rm{exotic}}$, which multiplies
the SM signal strength to predict the rate in the exotic model under test.
Both distributions show agreement between the background predictions and the
observed data over five orders of magnitude. No evidence for an excess of
exotic signal-like candidates is seen.

A number of systematic uncertainties among the various analyses affect
the sensitivity of the final result.  All correlations within and between
channels are taken into account in deriving the following combined limits,
cross sections, and $p$-values.  Uncertainties of
5\%~\cite{djouadibaglio,vhtheory} on the inclusive {\it WH} and {\it ZH}
production rates are estimated by varying the factorization and
renormalization scales.  We assign uncertainties to the Higgs boson
branching ratios as calculated in Ref.~\cite{denner11}.  Since the exotic
signals are normalized to the SM Higgs cross section, the same relative
uncertainties are assumed for the exotic production.  The largest sources
of uncertainty on the dominant backgrounds are the rates of
{\it V}+heavy-flavor jets.  The resulting uncertainties are up to 8\% of
the predicted values.  Because the various analyses use different methods
to obtain the $V$+heavy-flavor predictions, we treat their uncertainties
as uncorrelated between the $\ell\nu b{\bar{b}}$, the $\met b{\bar{b}}$,
and $\ell^+\ell^-b{\bar{b}}$ channels.  We use simulated events to study
the impact of the jet-energy-scale uncertainty~\cite{Bhatti:2005ai} on the
rates and shapes of the signal and background expectations.  We treat the
jet-energy-scale variations uncorrelated among the three analyses in the
combined search.  Uncertainties on lepton identification and trigger
efficiencies range from 2\% to 6\% and are applied to both signal and
MC-based background predictions.  The uncertainty on the integrated
luminosity is 6\%, of which 4.4\% originates from detector acceptance
uncertainties and 4.0\% is due to the uncertainty on the inelastic
$p{\bar{p}}$ cross section~\cite{inelppbarxs}.  The luminosity uncertainty
is correlated between the signal and MC-based background predictions.

\begin{figure}[htb]
 \begin{centering}
 \includegraphics[width=0.67\linewidth]{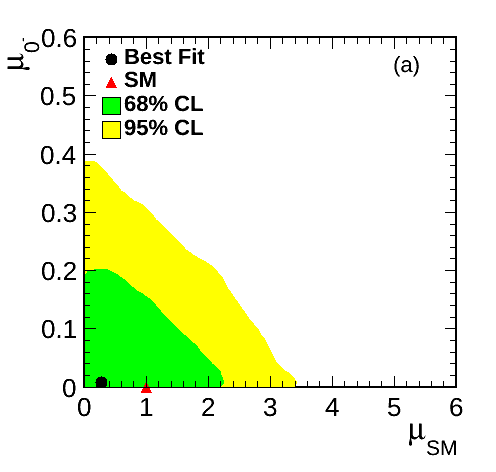}
 \includegraphics[width=0.67\linewidth]{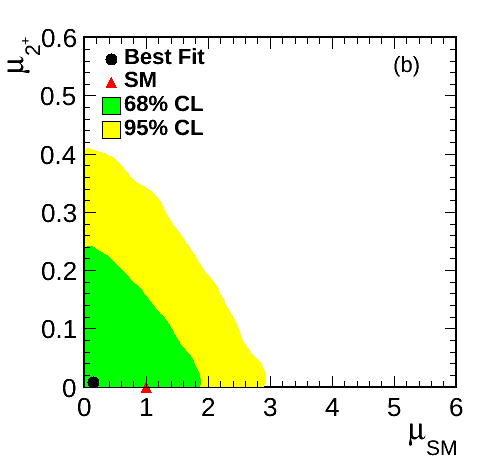}
 \caption{\label{fig:cdf2d}Combined two-dimensional posterior density of the
measured (a) $0^-$-vs.-$0^+$ and (b) $2^+$-vs.-$0^+$ cross sections, normalized
to the SM predictions.}
 \end{centering}
 \end{figure}

Bayesian exclusion limits at 95\% credibility level (C.L.)~\cite{pdgstats}
on the production rates times the branching fraction
${\cal B}(X\rightarrow b{\bar{b}})$ for $0^-$ and $2^+$ Higgs bosons are
reported in Table~\ref{tab:limits}, both separately for each channel and
combined, in units of the SM Higgs boson production rate.  The limits are
computed from a likelihood defined as the product of the probability
densities for the bin contents of the MVA histograms over all bins of each
histogram and all channel histograms, assuming Poisson probability densities
for the bin contents, uniform prior densities for the SM and exotic signal
strength modifiers $\mu_{\rm{exotic}}$ and $\mu_{\rm{SM}}$, and Gaussian prior
densities for the nuisance parameters describing systematic uncertainties. 
Posterior densities and upper limits on the SM and exotic Higgs boson rates
are obtained from pseudoexperiments (PEs), where in each PE the likelihood
is integrated over the nuisance parameters and then it is maximized.  The
medians of the distributions of results from PEs are used as the most probable
values.  The SM ratio between {\it WH} and {\it ZH} production rates is
assumed when combining {\it WX} and {\it ZX} searches.  Limits are listed
either assuming that the SM Higgs boson is present as a background, or absent.
Since the exotic $0^-$ and $2^+$ signals populate kinematic regions different
from those of the SM Higgs boson, and since the SM Higgs boson production
rate is small, the expected and observed limits on the exotic rates are very
similar whether the SM Higgs boson is present or not.  The observed combined
limits are somewhat stronger than expected, with an exclusion rate of
$\mu_{\rm{exotic}}<0.32$ in the $0^-$ case (approximately a one standard
deviation deficit), and $\mu_{\rm{exotic}}<0.35$ in the $2^+$ case (approximately
a two standard deviation deficit).  The $\met b{\bar{b}}$ channel carries
the largest weight in the combination.  A number of candidates somewhat
lower than expected appear in the most signal-like bins of the exotic
discriminants in this channel.  The two-dimensional cross section fits,
which allow for arbitrary rates of both SM and exotic Higgs bosons to be
simultaneously present, are shown in Fig.~\ref{fig:cdf2d}, separately for
the $0^-$ and $2^+$ searches.

 \begin{table}[htb]
 \begin{center}
 \caption{Limits at 95\% C.L. on $0^-$ and $2^+$ boson production assuming no
          SM Higgs boson background. In parentheses are the limits assuming
          SM Higgs boson background.}
 \label{tab:limits}
\begin{scriptsize}
 \begin{tabular}{ccccc}
 \hline\hline
         & \multicolumn{2}{c}{$0^-$} & \multicolumn{2}{c}{$2^+$} \\
 Channel & Obs  & Median exp & Obs  & Median exp \\
  & [limit/$H_{\rm SM}$] & [limit/$H_{\rm SM}$] & [limit/$H_{\rm SM}$] & [limit/$H_{\rm SM}$] \\
 \hline
$\ell\nu b{\bar{b}}$      & 0.59 (0.55) & 0.74 (0.78) & 1.05 (0.99) & 1.01 (1.03) \\
$\ell^+\ell^- b{\bar{b}}$ & 1.86 (1.77) & 1.46 (1.52) & 1.57 (1.49) & 1.59 (1.61) \\
$\met b{\bar{b}}$         & 0.49 (0.43) & 0.68 (0.69) & 0.41 (0.37) & 0.79 (0.83) \\
Combined                  & 0.32 (0.28) & 0.44 (0.45) & 0.35 (0.31) & 0.54 (0.56) \\
 \hline\hline
 \end{tabular}
\end{scriptsize}
 \end{center}
 \end{table}

We report the observed values and the expected distributions of the
log-likelihood ratio (LLR)~\cite{pdgstats} in the SM and the exotic hypotheses
and list the combined results in Table~\ref{tab:p-values}.  The Table includes
the $p$-values for the null and test hypotheses, defined as the conditional
probabilities $p_{\rm null}=P({\rm LLR\le LLR_{obs}\vert SM})$ and
$p_{\rm test}=P({\rm LLR\ge LLR_{obs}\vert exotic})$, respectively, the values
of ${\rm CL}_s=p_{\rm test}/(1-p_{\rm null})$, and the equivalent number of
Gaussian standard deviations $z$ corresponding to each $p$-value, defined
by $p=[1-{\rm erf}(z/\sqrt{2})]/2$~\cite{pdgstats}.  There is a deficit in the
observed number of events in the signal-like bins of the exotic discriminant,
which is visible in Fig.~\ref{fig:sb} in both the $0^-$ and the $2^+$ searches.
The dominant contribution to this deficit comes from the
$WX+ZX\rightarrow\met b{\bar{b}}$ search.  This deficit in the exotic search
is not evidence against the SM Higgs boson, as the exotic search tests for
events with different kinematic properties (high $M_{Vb{\bar{b}}}$) than those
of the SM Higgs boson.  Indeed, the combined cross section fit, shown in
Fig.~\ref{fig:cdf2d}, is consistent with the SM Higgs boson rate with a
discrepancy of less than 0.5 standard deviations.

\begin{table*}[htb]
\begin{center}
\caption{LLR and $p$-values for the test hypotheses.  The SM hypothesis
includes a SM Higgs boson.  The significances corresponding to the $p$-values
and CL$_{\rm{s}}$ are given in parentheses in units of standard deviation
(s.d.).  The negative signal significance $p_{\rm null}$ reflects the deficit
of signal-like events compared with the background prediction.}
\label{tab:p-values}
\begin{tabular}{lll}
\hline\hline
                                              & ~~~$0^-$                           & ~~~$2^+$                          \\\hline
LLR$_{\rm obs}$                                & ~~24.6                             &  ~~20.8                           \\
LLR$_{\rm SM,~median}$                          & ~~13.2                             &  ~~~9.5                           \\
LLR$_{\rm exotic,~median}$                      & $-$15.5                             & $-$10.8                           \\
$p_{\rm null}$                                &  0.943~             ($-$1.58 s.d.)   &  0.967~        ($-$1.83 s.d.)     \\
Median expected p$_{\rm null}$ (if exotic)~~~ &  3.87$\times 10^{-4}$~ (3.95 s.d.)~~~ &  4.96$\times 10^{-4}$~ (3.29 s.d.) \\
$p_{\rm test}$                                &  1.72$\times 10^{-7}$~ (5.10 s.d.)    &  7.65$\times 10^{-7}$~ (4.81 s.d.) \\
Median expected p$_{\rm test}$ (if SM)        &  1.36$\times 10^{-4}$~ (3.64 s.d.)    &  1.01$\times 10^{-3}$~ (3.09 s.d.) \\
CL$_{\rm s}$                                  &  3.03$\times 10^{-5}$~ (4.52 s.d.)    &  2.29$\times 10^{-4}$~ (4.08 s.d.) \\
Median expected CL$_{\rm s}$                  &  2.72$\times 10^{-4}$~ (3.46 s.d.)    &  2.01$\times 10^{-3}$~ (2.88 s.d.) \\\hline\hline
\end{tabular}
\end{center}
\end{table*}


In conclusion, we search in the entire CDF data sample for Higgs-boson-like
particles of the same mass, production and decay modes, and production rates
as the discovered SM Higgs boson candidate, but with $0^-$ or $2^+$ spin-parity
quantum numbers.  We observe no significant deviations from the SM predictions
with a Higgs boson of mass $m_H\approx 125$~GeV/$c^2$, and set bounds on the
possible rate of production of $0^-$ and $2^+$ exotic states, both allowing
for an admixture of SM production and exotic production, and assuming only
exotic production.

\begin{center}
{\bf Acknowledgments}
\end{center}

We thank the Fermilab staff and the technical staffs of the
participating institutions for their vital contributions. This work
was supported by the U.S. Department of Energy and National Science
Foundation; the Italian Istituto Nazionale di Fisica Nucleare; the
Ministry of Education, Culture, Sports, Science and Technology of
Japan; the Natural Sciences and Engineering Research Council of
Canada; the National Science Council of the Republic of China; the
Swiss National Science Foundation; the A.P. Sloan Foundation; the
Bundesministerium f\"ur Bildung und Forschung, Germany; the Korean
World Class University Program, the National Research Foundation of
Korea; the Science and Technology Facilities Council and the Royal
Society, United Kingdom; the Russian Foundation for Basic Research;
the Ministerio de Ciencia e Innovaci\'{o}n, and Programa
Consolider-Ingenio 2010, Spain; the Slovak R\&D Agency; the Academy
of Finland; the Australian Research Council (ARC); and the EU community
Marie Curie Fellowship Contract No. 302103.

 \begin{figure*}
 \begin{centering}
{\bf Supplemental material}

\vspace{2cm}
 \includegraphics[width=0.45\linewidth]{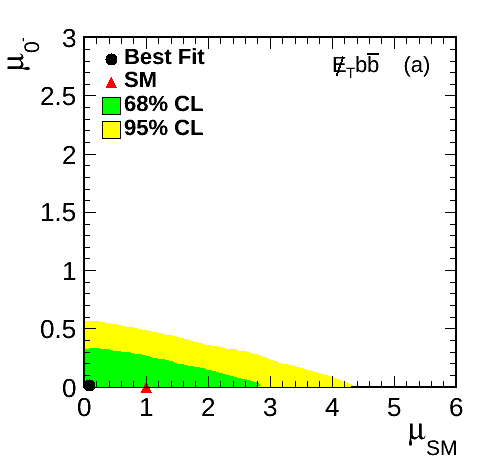}
 \includegraphics[width=0.45\linewidth]{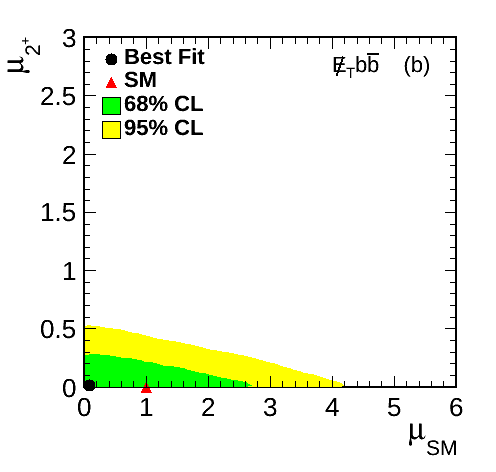}
 \label{fig:cdfmetbb}
 \caption{$\met b\bar{b}$ two-dimensional posterior density of the measured
(a) $0^-$-vs.-$0^+$ and (b) $2^+$-vs.-$0^+$ cross sections, normalized to the
SM predictions.}
 \end{centering}
 \end{figure*}

\begin{figure*}[htb]
 \begin{centering}
 \includegraphics[width=0.45\linewidth]{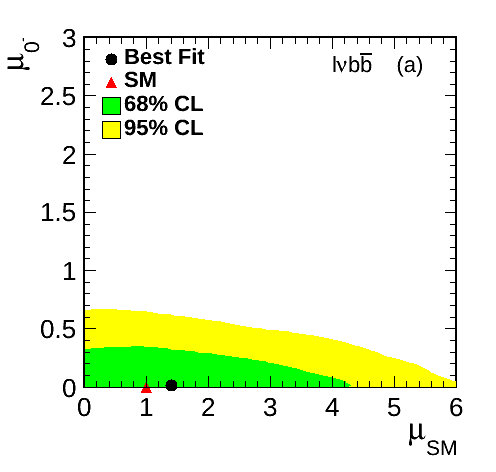}
 \includegraphics[width=0.45\linewidth]{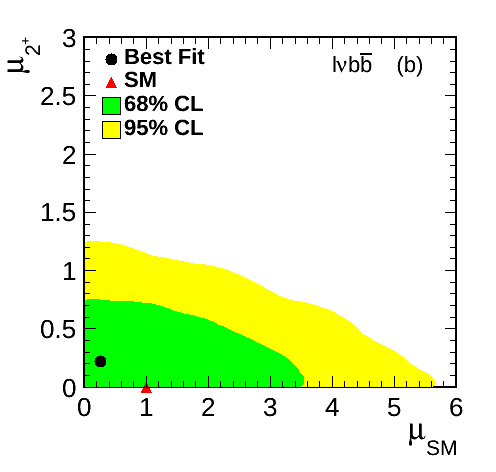}
 \label{fig:cdflvbb}
 \caption{$\ell\nu b\bar{b}$ two-dimensional posterior density of the measured
(a) $0^-$-vs.-$0^+$ and (b) $2^+$-vs.-$0^+$ cross sections, normalized to the
SM predictions.}
 \end{centering}
 \end{figure*}

\begin{figure*}[htb]
 \begin{centering}
 \includegraphics[width=0.45\linewidth]{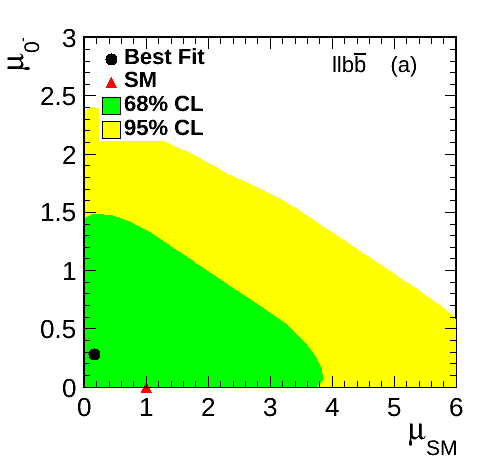}
 \includegraphics[width=0.45\linewidth]{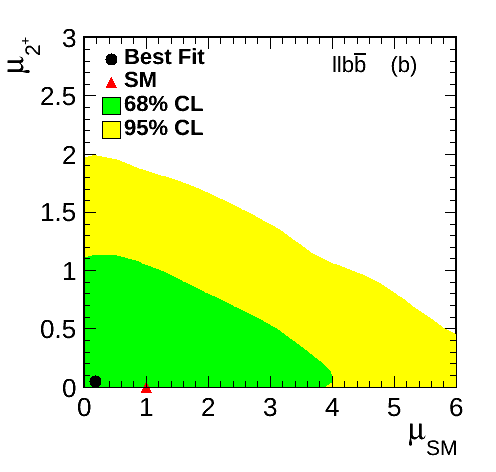}
 \label{fig:cdfllbb}
 \caption{
$\ell\ell b\bar{b}$ two-dimensional posterior density of the measured (a)
$0^-$-vs.-$0^+$ and (b) $2^+$-vs.-$0^+$ cross sections, normalized to the SM
predictions.}
 \end{centering}
 \end{figure*}

 \begin{figure*}[htb]
 \label{fig:cdfmetbb_p}
 \begin{centering}
 \includegraphics[width=0.45\linewidth]{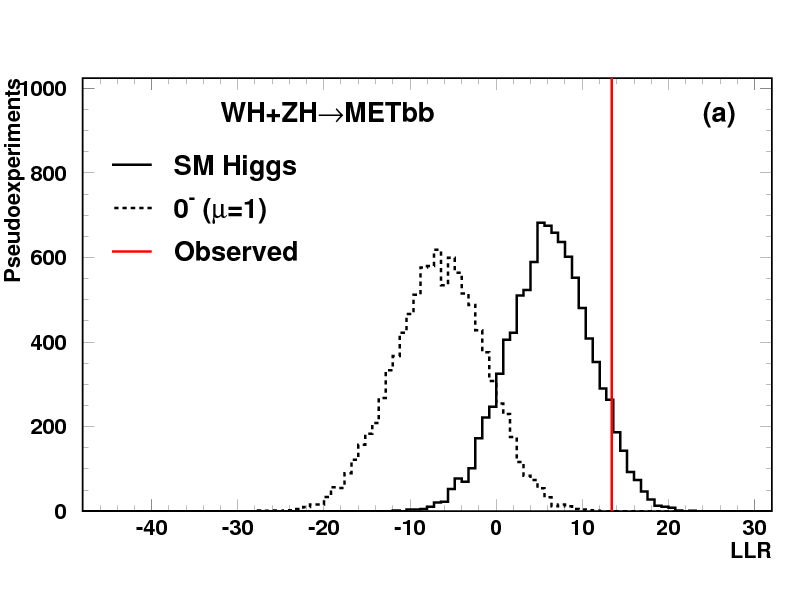}
 \includegraphics[width=0.45\linewidth]{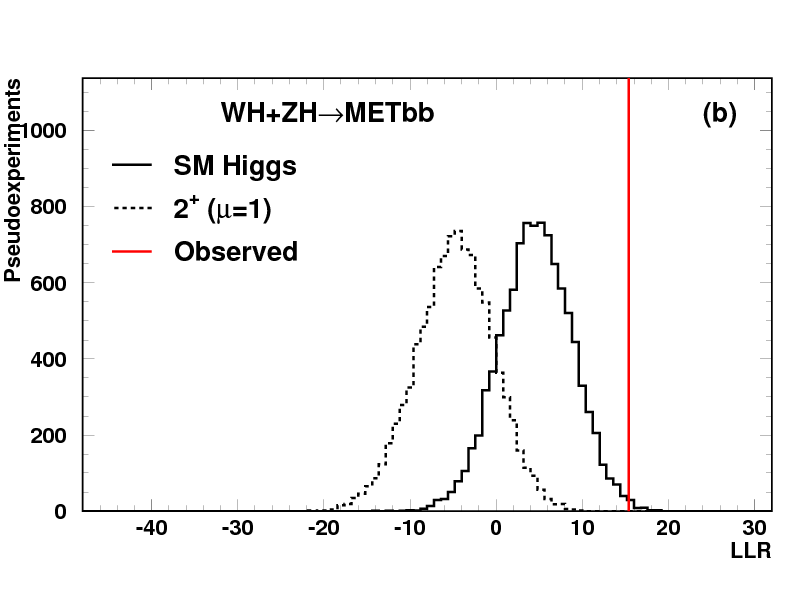}
 \caption{$\met b\bar{b}$ LLR distributions for (a) $0^-$ and (b) $2^+$
hypotheses, assuming $\mu$ = 1.
}
 \end{centering}
 \end{figure*}

\begin{figure*}[htb]
 \label{fig:cdflvbb_p}
 \begin{centering}
 \includegraphics[width=0.45\linewidth]{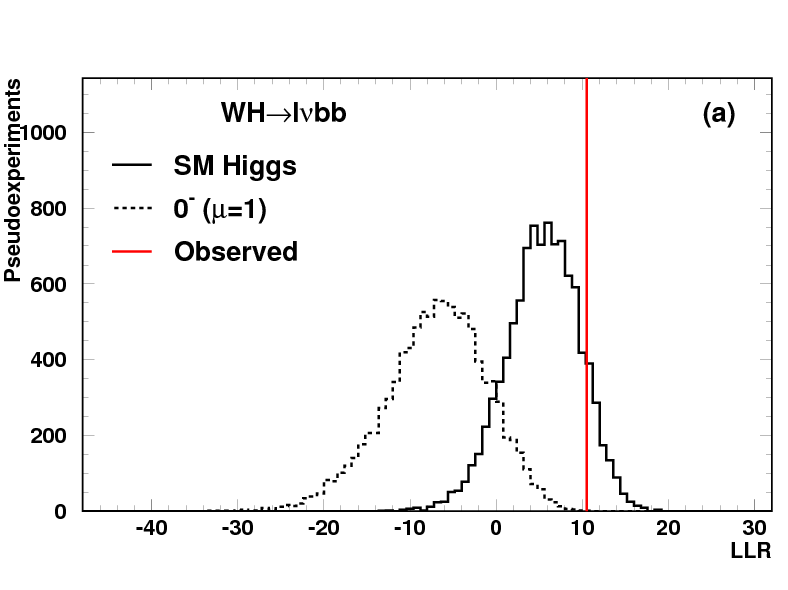}
 \includegraphics[width=0.45\linewidth]{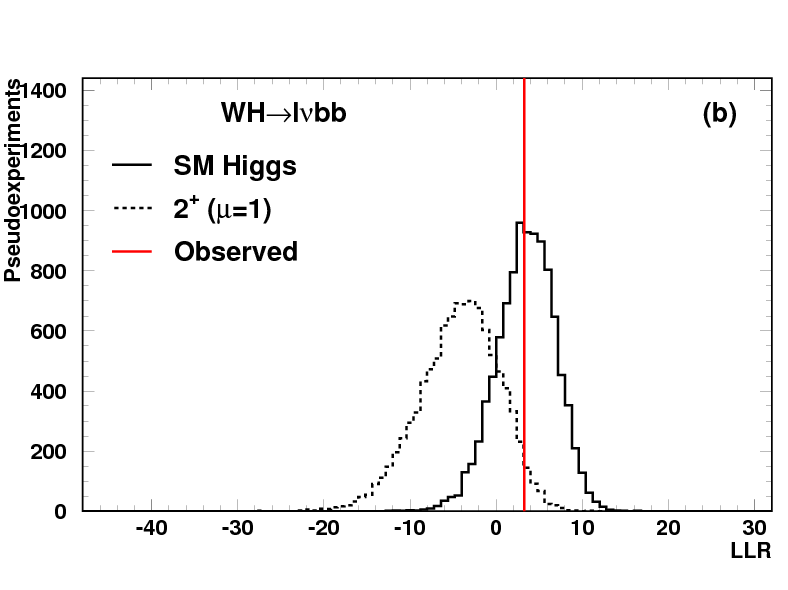}
 \caption{$\ell\nu b\bar{b}$ LLR distributions for (a) $0^-$ and (b) $2^+$
hypotheses, assuming $\mu$ = 1.
}
 \end{centering}
 \end{figure*}

\begin{figure*}[htb]
 \label{fig:cdfllbb_p}
 \begin{centering}
 \includegraphics[width=0.45\linewidth]{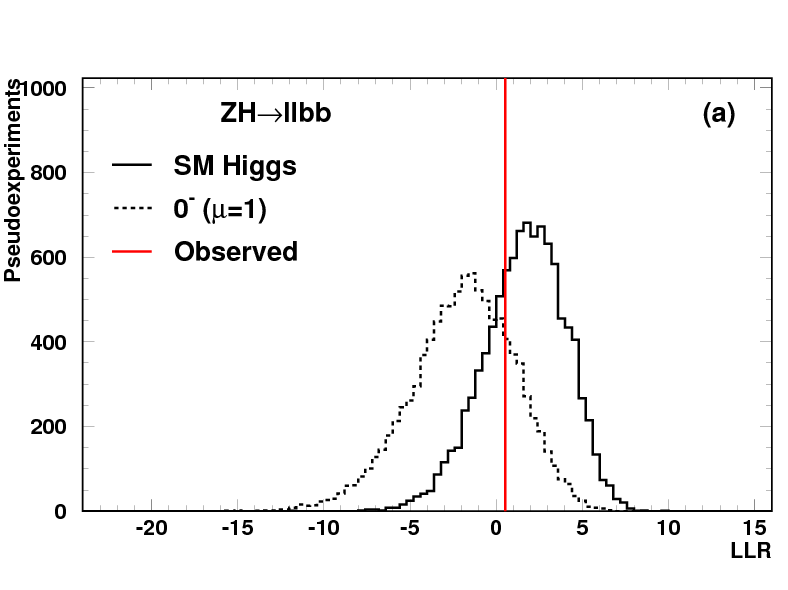}
 \includegraphics[width=0.45\linewidth]{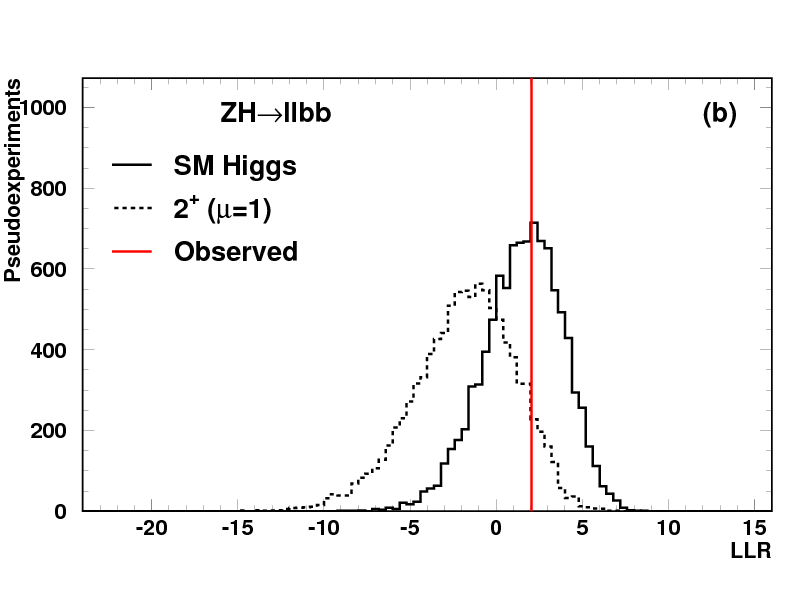}
 \caption{$\ell\ell b\bar{b}$ LLR distributions for (a) $0^-$ and (b) $2^+$
hypotheses, assuming $\mu$ = 1.
}
 \end{centering}
 \end{figure*}

\begin{figure*}[htb]
 \label{fig:cdf_p}
 \begin{centering}
 \includegraphics[width=0.45\linewidth]{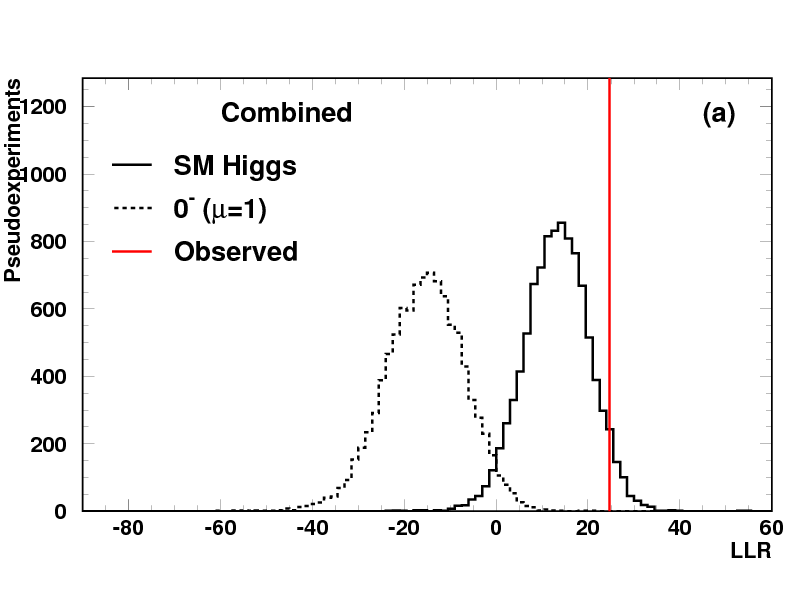}
 \includegraphics[width=0.45\linewidth]{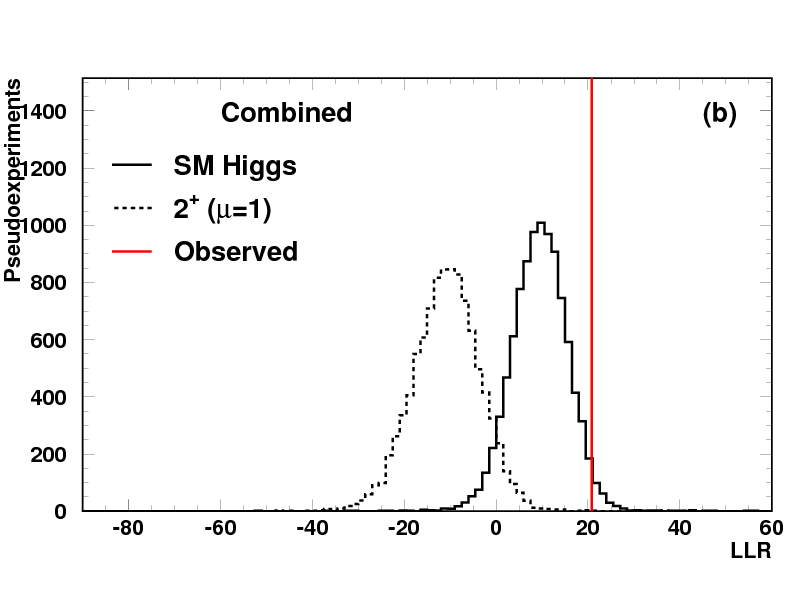}
 \caption{Combined LLR distributions for (a) $0^-$ and (b) $2^+$
hypotheses, assuming $\mu$ = 1.
}
 \end{centering}
 \end{figure*}

\begin{figure*}[htb]
 \begin{centering}
 \includegraphics[width=0.45\linewidth]{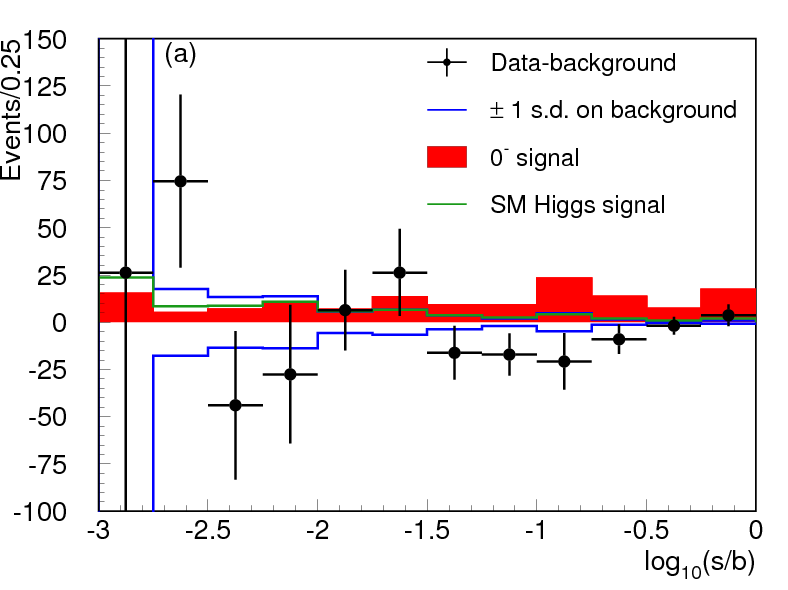}
 \includegraphics[width=0.45\linewidth]{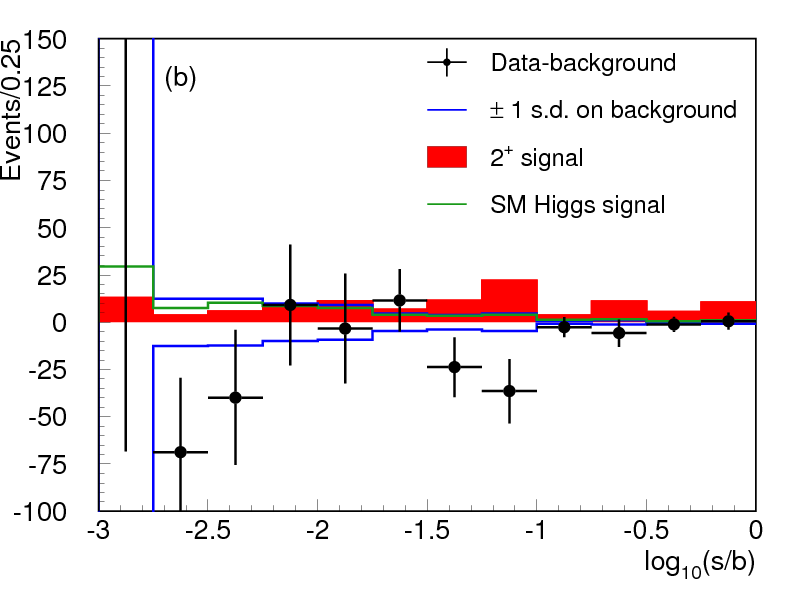}
 \label{fig:bgsub}
 \caption{Background-subtracted distribution of the discriminant histograms,
summed for bins with similar signal-to-background ratio ($s/b$) over all
contributing Higgs boson search channels, for $m_H=125$~GeV/$c^2$, for (a)
the $0^-$ search and (b) the $2^+$ search.  The background is fit to the data
in each case, and the uncertainty on the background, shown with dashed lines,
is after the fit.  The exotic signal model, scaled to the SM Higgs boson
expectation, is shown with a filled histogram.  The SM Higgs boson expectation
is also shown with a solid line.  The error bars shown on the data points
correspond in each bin to the square root of the sum of the expected signal
and background yields.  Underflows and overflows are collected into the
leftmost and rightmost bins, respectively.}
 \end{centering}
 \end{figure*}

\begin{figure*}[htb]
 \begin{centering}
\includegraphics[width=0.45\linewidth]{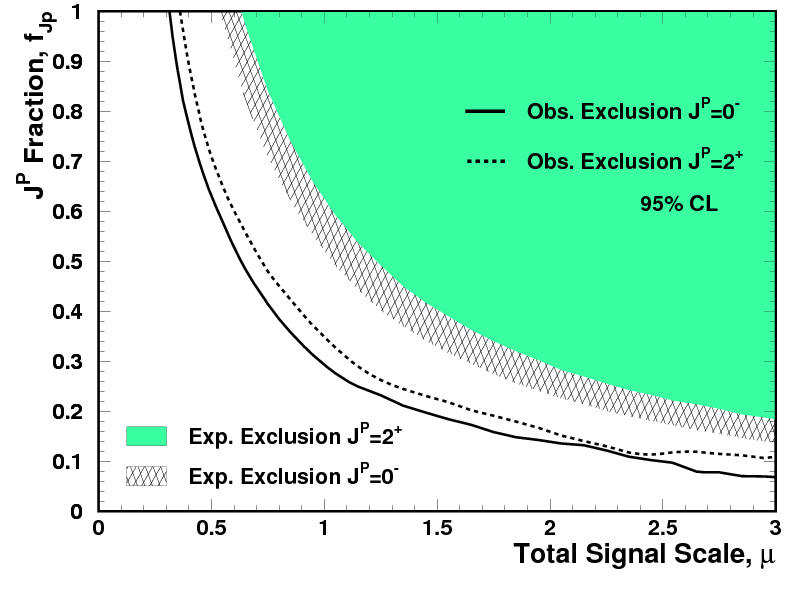}
 \label{fig:fraclim}
 \caption{Exclusion limits on the fraction of exotic boson in the total signal
for the exotic $+$ SM Higgs boson admixture hypotheses. The limits are shown as
functions of the total signal production rate relative to the SM prediction for
the Higgs boson rate.}
 \end{centering}
 \end{figure*}

\clearpage

\begin{figure*}[htb]
 \begin{centering}
\includegraphics[width=0.45\linewidth]{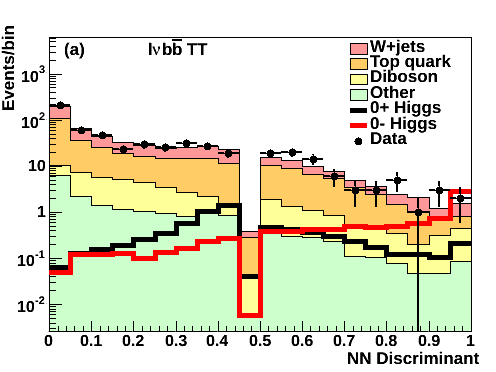}
 \includegraphics[width=0.45\linewidth]{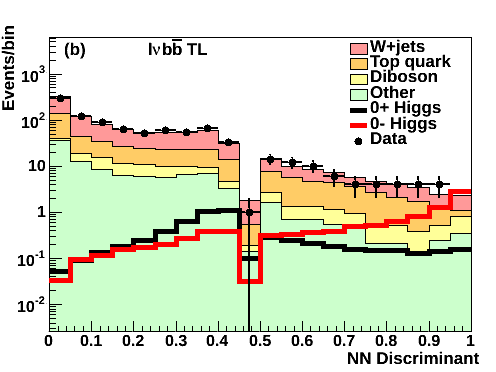}
 \caption{Discriminant function for (a) the tight-tight and (b) the
tight-loose $b$-jet categories of the $\ell\nu b{\bar{b}}$ search channel
for a 0$^-$ state.  Events are first categorized according to their exotic
discriminant outputs.  If these pass below a threshold of 0.5, then the SM
discriminant function is used instead, scaled to the range [0,0.5].  In this
way, kinematic regions that are the most sensitive to testing for the presence
of an exotic Higgs boson are used first, and non-exotic-like events are used
to compare the data with the SM Higgs boson hypothesis.}
 \end{centering}
 \end{figure*}

\begin{figure*}[htb]
 \begin{centering}
\includegraphics[width=0.45\linewidth]{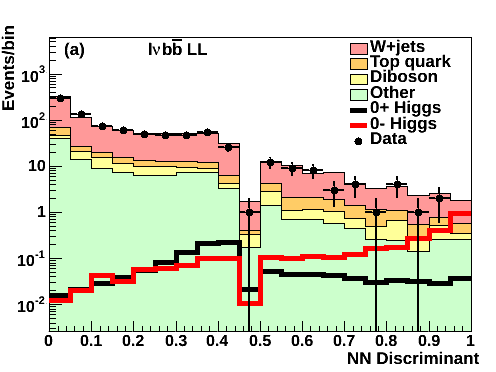}
 \includegraphics[width=0.45\linewidth]{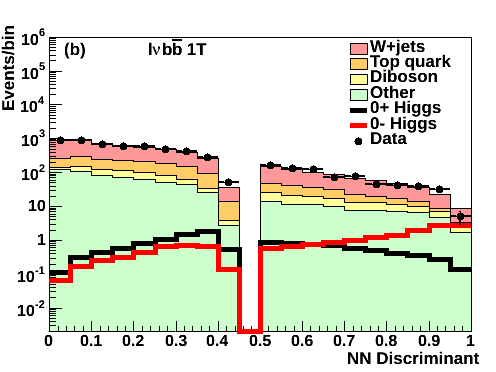}
 \caption{Discriminant function for (a) the loose-loose and
          (b) the 1-tight (right) $b$-jet categories of the
          $\ell\nu b{\bar{b}}$ search channel for a 0$^-$ state.}
 \end{centering}
 \end{figure*}

\begin{figure*}[htb]
 \begin{centering}
\includegraphics[width=0.45\linewidth]{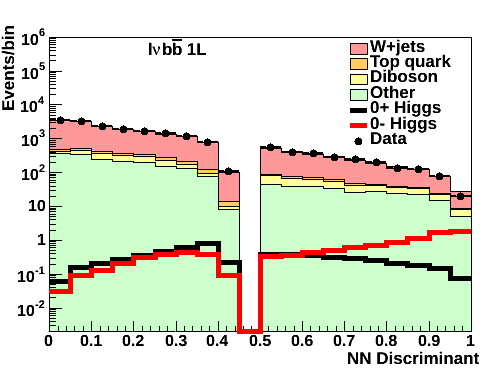}
 \caption{Discriminant function for the 1-loose $b$-jet
          category of the $\ell\nu b{\bar{b}}$ search channel
          for a 2$^+$ state.}
 \end{centering}
 \end{figure*}

\begin{figure*}[htb]
 \begin{centering}
\includegraphics[width=0.45\linewidth]{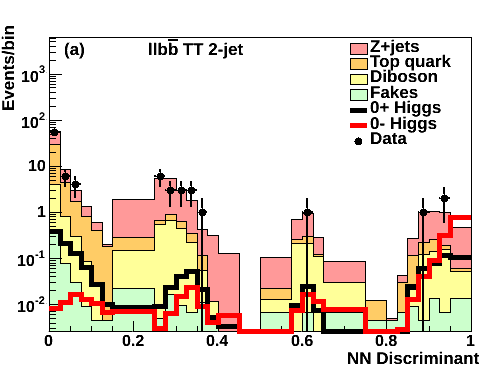}
\includegraphics[width=0.45\linewidth]{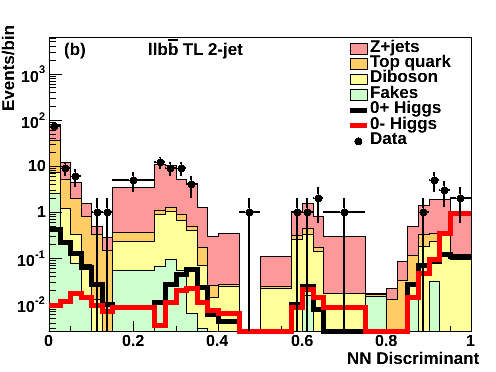}
 \caption{Discriminant function for (a) the tight-tight and
          (b) the tight-loose $b$-jet categories in the 2-jet bin
          of the $\ell\ell b{\bar{b}}$ search channel for a 0$^-$ state.}
 \end{centering}
 \end{figure*}

\begin{figure*}[htb]
 \begin{centering}
\includegraphics[width=0.45\linewidth]{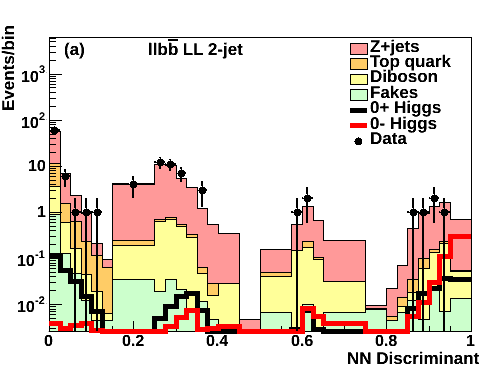}
\includegraphics[width=0.45\linewidth]{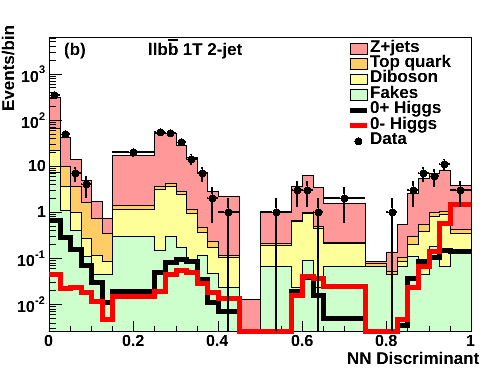}
 \caption{Discriminant function for (a) the loose-loose and (b)
          the 1-tight $b$-jet categories in the 2-jet bin of the
          $\ell\ell b{\bar{b}}$ search channel for a 0$^-$ state.}
 \end{centering}
 \end{figure*}

\begin{figure*}[htb]
 \begin{centering}
\includegraphics[width=0.45\linewidth]{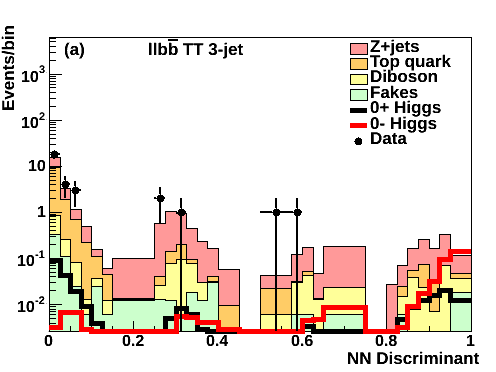}
\includegraphics[width=0.45\linewidth]{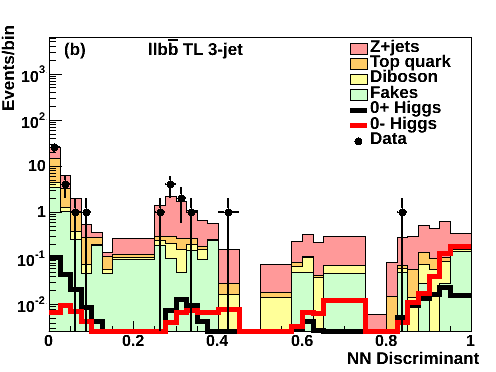}
 \caption{Discriminant function for (a) the tight-tight and (b)
          the tight-loose $b$-jet categories in the 3-jet bin of the
          $\ell\ell b{\bar{b}}$ search channel for a 0$^-$ state.}
 \end{centering}
 \end{figure*}

\begin{figure*}[htb]
 \begin{centering}
\includegraphics[width=0.45\linewidth]{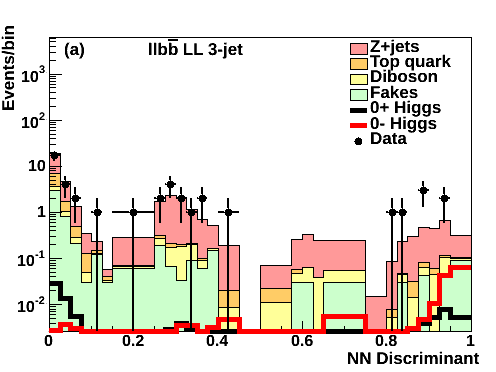}
\includegraphics[width=0.45\linewidth]{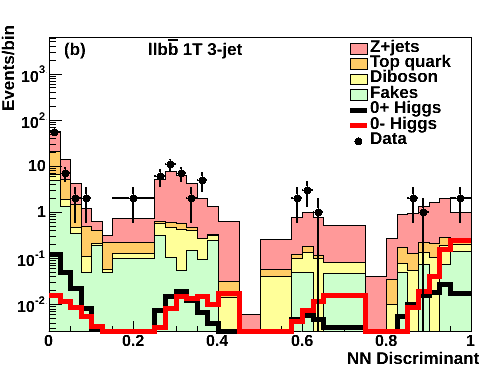}
 \caption{Discriminant function for (a) the loose-loose and (b)
          the 1-tight $b$-jet categories in the 3-jet bin of the
          $\ell\ell b{\bar{b}}$ search channel for a 0$^-$ state.}
 \end{centering}
 \end{figure*}

\begin{figure*}[htb]
 \begin{centering}
\includegraphics[width=0.45\linewidth]{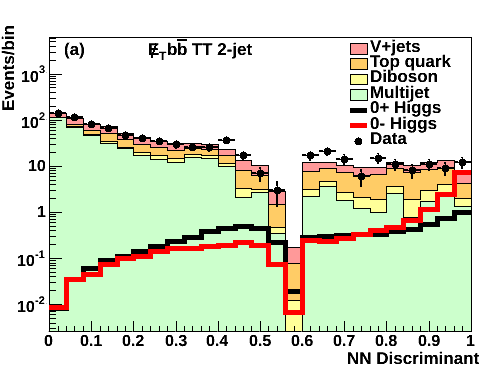}
\includegraphics[width=0.45\linewidth]{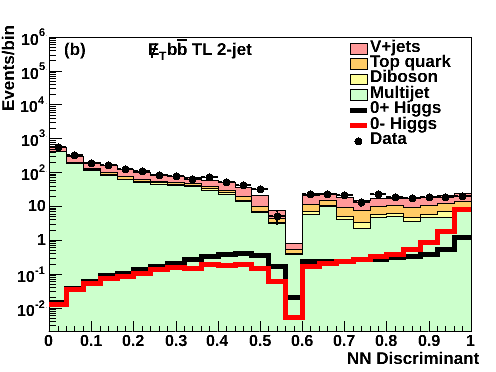}
 \caption{Discriminant function for (a) the tight-tight and (b)
the tight-loose $b$-jet categories in the 2-jet bin of the $\met b{\bar{b}}$
search channel for a 0$^-$ state.  Events are first categorized according to
their exotic discriminant outputs.  If these pass below a threshold (typically
0.5 on a scale of 0 to 1), then the SM discriminant function is used instead,
scaled to the range [0,0.5].  In this way, kinematic regions that are the most
sensitive to testing for the presence of an exotic Higgs boson are used first,
and non-exotic-like events are used to compare the data with the SM Higgs
boson hypothesis.}
 \end{centering}
 \end{figure*}

\begin{figure*}[htb]
 \begin{centering}
\includegraphics[width=0.45\linewidth]{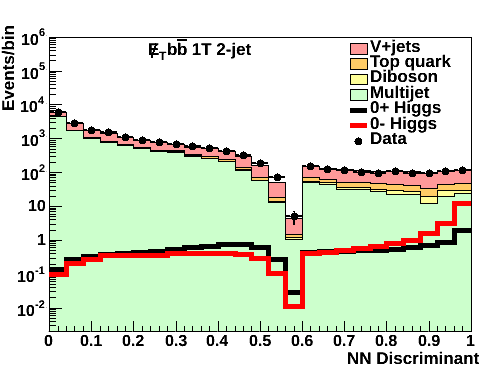}
 \caption{Discriminant function for the 1-tight $b$-jet category
          in the 2-jet bin of the $\met b{\bar{b}}$ search channel
          for a 0$^-$ state.}
 \end{centering}
 \end{figure*}

\begin{figure*}[htb]
 \begin{centering}
\includegraphics[width=0.45\linewidth]{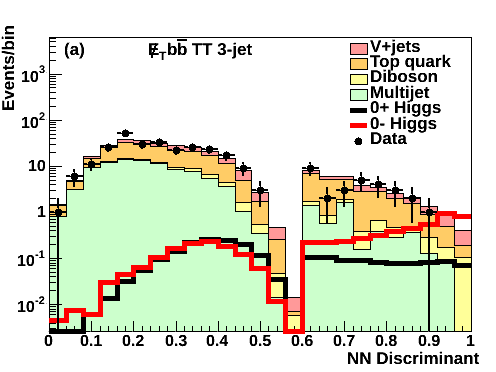}
\includegraphics[width=0.45\linewidth]{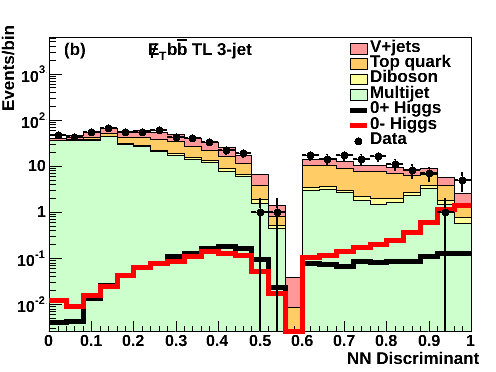}
 \caption{Discriminant function for (a) the tight-tight and (b)
          the tight-loose $b$-jet categories in the 3-jet bin of the
          $\met b{\bar{b}}$ search channel for a 0$^-$ state.}
 \end{centering}
 \end{figure*}

\begin{figure*}[htb]
 \begin{centering}
\includegraphics[width=0.45\linewidth]{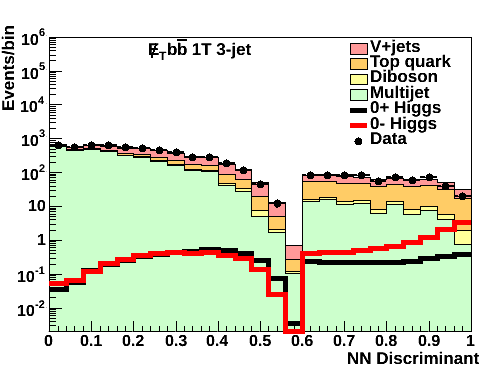}
 \caption{Discriminant function for the 1-tight $b$-jet category
          in the 3-jet bin of the $\met b{\bar{b}}$ search channel
          for a 0$^-$ state.}
 \end{centering}
 \end{figure*}

\clearpage

\begin{figure*}[htb]
 \begin{centering}
\includegraphics[width=0.45\linewidth]{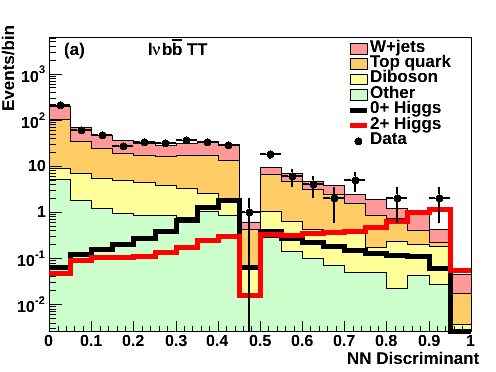}
 \includegraphics[width=0.45\linewidth]{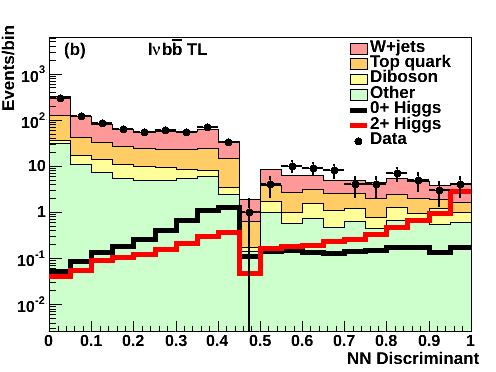}
 \caption{Discriminant function for (a) the tight-tight and (b) the
tight-loose $b$-jet categories of the $\ell\nu b{\bar{b}}$ search channel
for a 2$^+$ state.  Events are first categorized according to their exotic
discriminant outputs.  If these pass below a threshold of 0.5, then the SM
discriminant function is used instead, scaled to the range [0,0.5].  In this
way, kinematic regions that are the most sensitive to testing for the presence
of an exotic Higgs boson are used first, and non-exotic-like events are used
to compare the data with the SM Higgs boson hypothesis.}
 \end{centering}
 \end{figure*}

\begin{figure*}[htb]
 \begin{centering}
\includegraphics[width=0.45\linewidth]{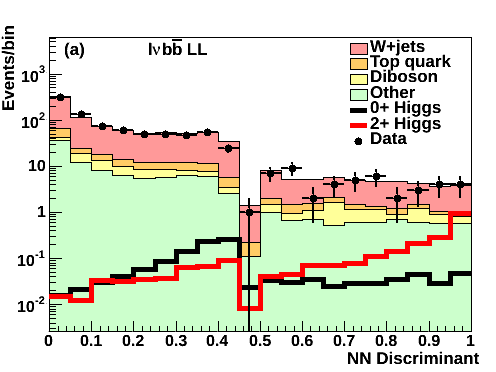}
 \includegraphics[width=0.45\linewidth]{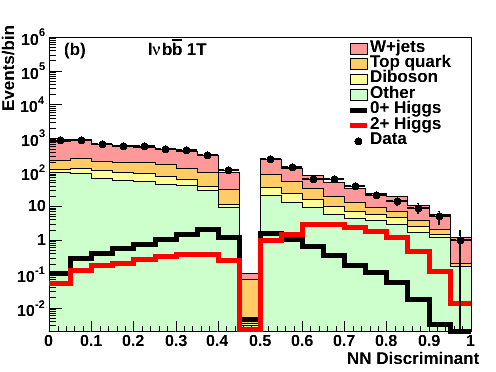}
 \caption{Discriminant function for (a) the loose-loose and
          (b) the 1-tight $b$-jet categories of the $\ell\nu b{\bar{b}}$
          search channel for a 2$^+$ state.}
 \end{centering}
 \end{figure*}

\begin{figure*}[htb]
 \begin{centering}
\includegraphics[width=0.45\linewidth]{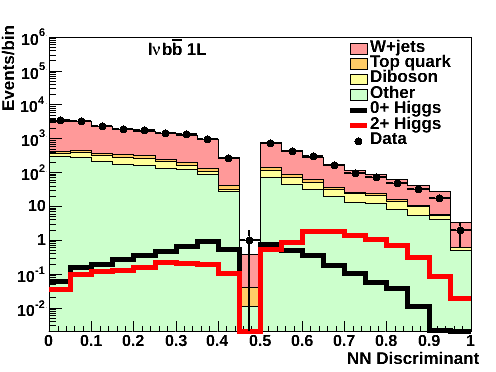}
 \caption{Discriminant function for the 1-loose $b$-jet
          category of the $\ell\nu b{\bar{b}}$ search channel
          for a 2$^+$ state.}
 \end{centering}
 \end{figure*}

\begin{figure*}[htb]
 \begin{centering}
\includegraphics[width=0.45\linewidth]{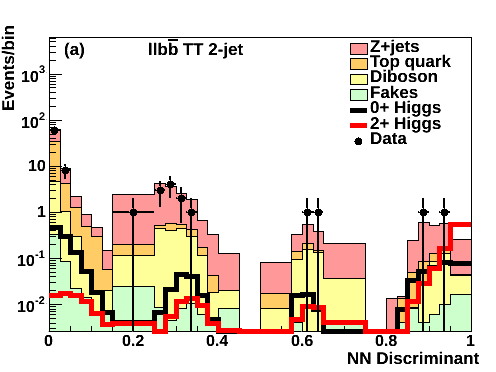}
\includegraphics[width=0.45\linewidth]{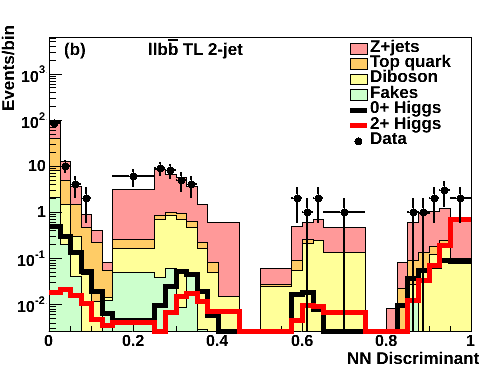}
 \caption{Discriminant function for (a) the tight-tight and
          (b) the tight-loose $b$-jet categories in the 2-jet bin
          of the $\ell\ell b{\bar{b}}$ search channel for a 2$^+$ state.}
 \end{centering}
 \end{figure*}

\begin{figure*}[htb]
 \begin{centering}
\includegraphics[width=0.45\linewidth]{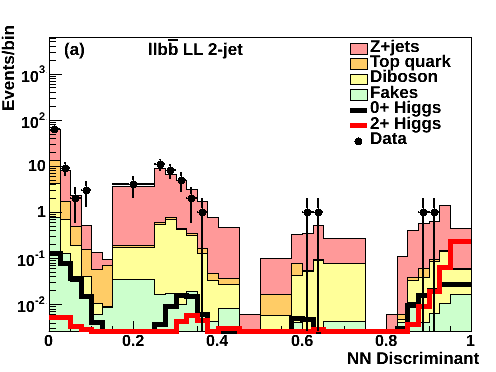}
\includegraphics[width=0.45\linewidth]{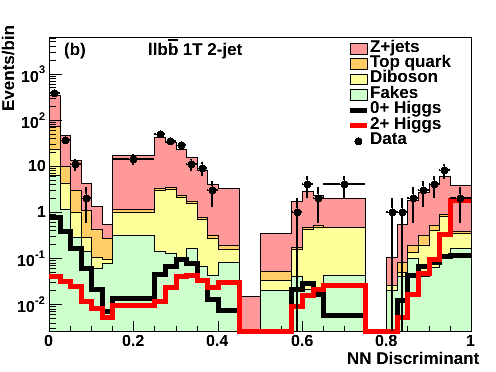}
 \caption{Discriminant function for (a) the loose-loose and
          (b) the 1-tight $b$-jet categories in the 2-jet bin of
          the $\ell\ell b{\bar{b}}$ search channel for a 2$^+$ state.}
 \end{centering}
 \end{figure*}

\begin{figure*}[htb]
 \begin{centering}
\includegraphics[width=0.45\linewidth]{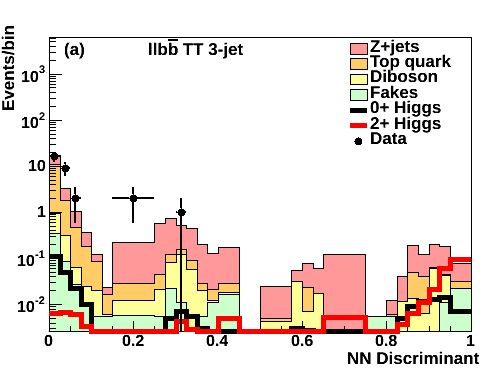}
\includegraphics[width=0.45\linewidth]{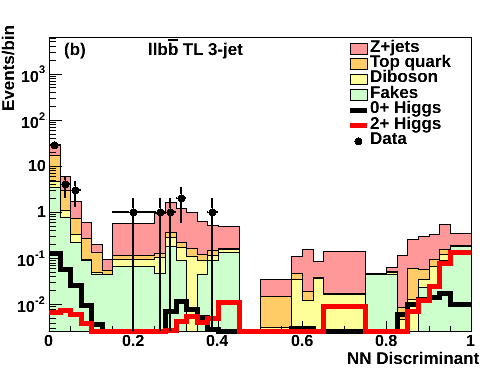}
 \caption{Discriminant function for (a) the tight-tight and
          (b) the tight-loose $b$-jet categories in the 3-jet bin
          of the $\ell\ell b{\bar{b}}$ search channel for a 2$^+$ state.}
 \end{centering}
 \end{figure*}

\begin{figure*}[htb]
 \begin{centering}
\includegraphics[width=0.45\linewidth]{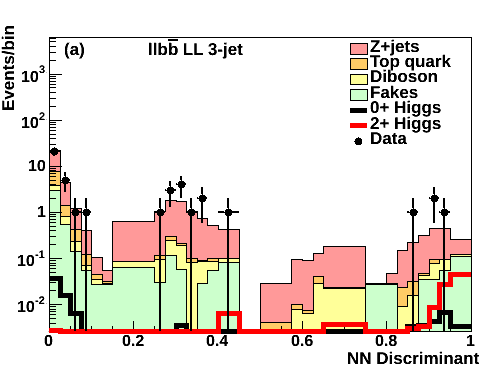}
\includegraphics[width=0.45\linewidth]{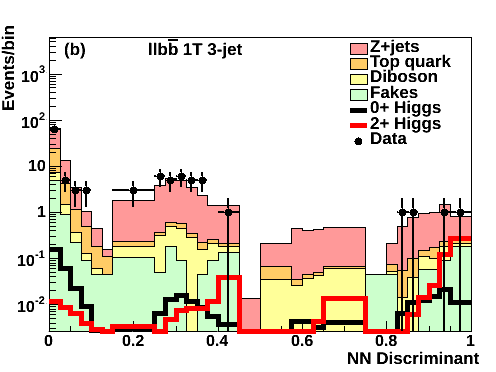}
 \caption{Discriminant function for (a) the loose-loose and
          (b) the 1-tight (right) $b$-jet categories in the 3-jet
          bin of the $\ell\ell b{\bar{b}}$ search channel for a 2$^+$ state.}
 \end{centering}
 \end{figure*}

\begin{figure*}[htb]
 \begin{centering}
\includegraphics[width=0.45\linewidth]{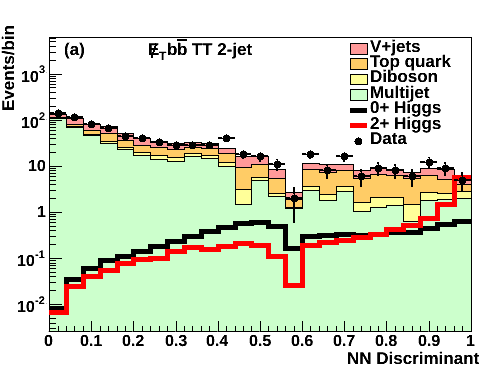}
\includegraphics[width=0.45\linewidth]{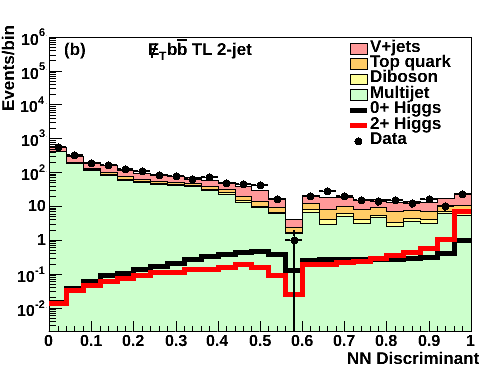}
 \caption{Discriminant function for (a) the tight-tight and
(b) the tight-loose $b$-jet categories in the 2-jet bin of the
$\met b{\bar{b}}$ search channel for a 2$^+$ state.  Events are first
categorized according to their exotic discriminant outputs.  If these pass
below a threshold of 0.6, then the SM discriminant function is used instead,
scaled to the range [0,0.5].  In this way, kinematic regions that are the
most sensitive to testing for the presence of an exotic Higgs boson are used
first, and non-exotic-like events are used to compare the data with the SM
Higgs boson hypothesis.}
 \end{centering}
 \end{figure*}

\begin{figure*}[htb]
 \begin{centering}
\includegraphics[width=0.45\linewidth]{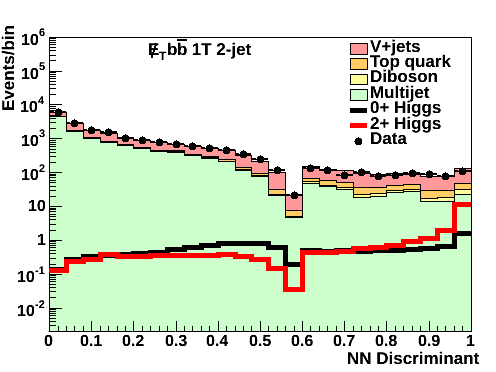}
 \caption{Discriminant function for the 1-tight $b$-jet category
          in the 2-jet bin of the $\met b{\bar{b}}$ search channel
          for a 2$^+$ state.}
 \end{centering}
 \end{figure*}

\begin{figure*}[htb]
 \begin{centering}
\includegraphics[width=0.45\linewidth]{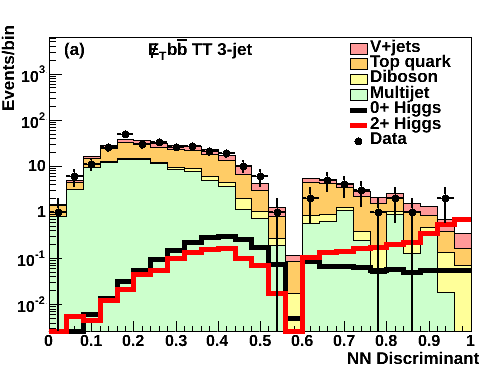}
\includegraphics[width=0.45\linewidth]{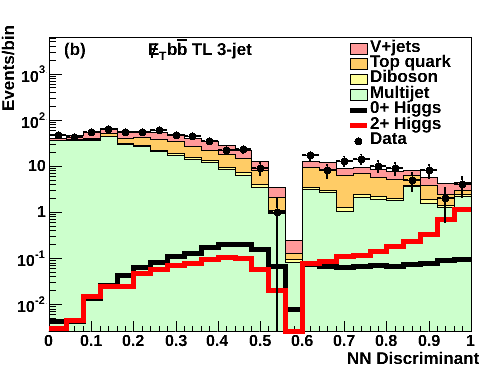}
 \caption{Discriminant function for (a) the tight-tight and
          (b) the tight-loose $b$-jet categories in the 3-jet bin
          of the $\met b{\bar{b}}$ search channel for a 2$^+$ state.}
 \end{centering}
 \end{figure*}

\begin{figure*}[htb]
 \begin{centering}
\includegraphics[width=0.45\linewidth]{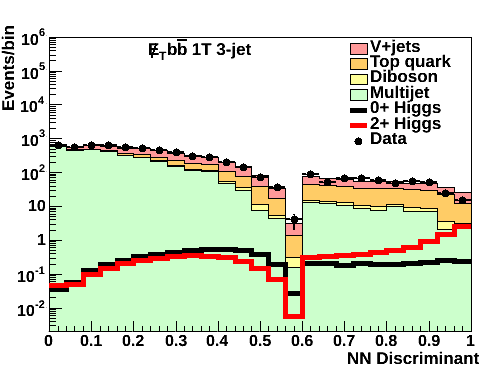}
 \caption{Discriminant function for the 1-tight $b$-jet category
          in the 3-jet bin of the $\met b{\bar{b}}$ search channel
          for a 2$^+$ state.}
 \end{centering}
 \end{figure*}

\clearpage

\begin{figure*}[htb]
 \begin{centering}
\includegraphics[width=0.45\linewidth]{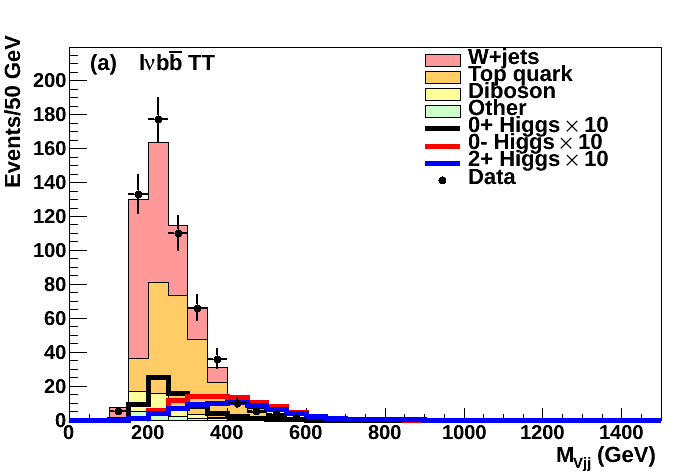}
 \includegraphics[width=0.45\linewidth]{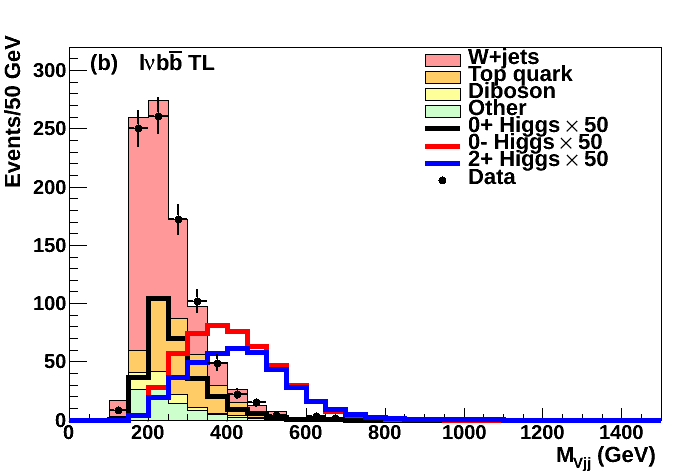}
 \caption{Reconstructed $Vb\bar b$ transverse invariant mass distribution
          for (a) the tight-tight and (b) the tight-loose $b$-jet categories
          of the $\ell\nu b{\bar{b}}$ search channel.}
 \end{centering}
 \end{figure*}

\begin{figure*}[htb]
 \begin{centering}
\includegraphics[width=0.45\linewidth]{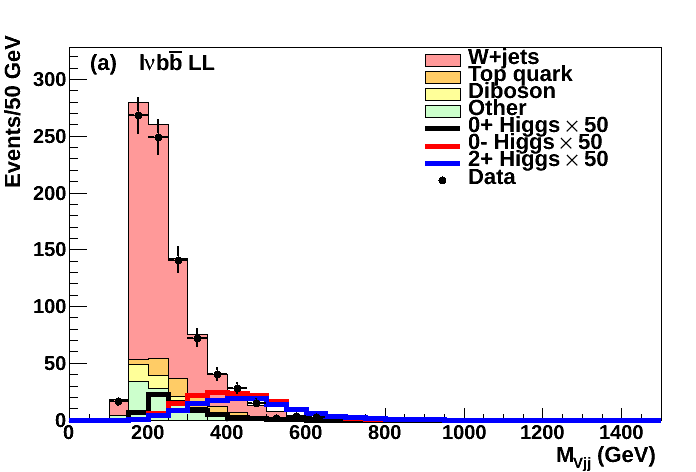}
 \includegraphics[width=0.45\linewidth]{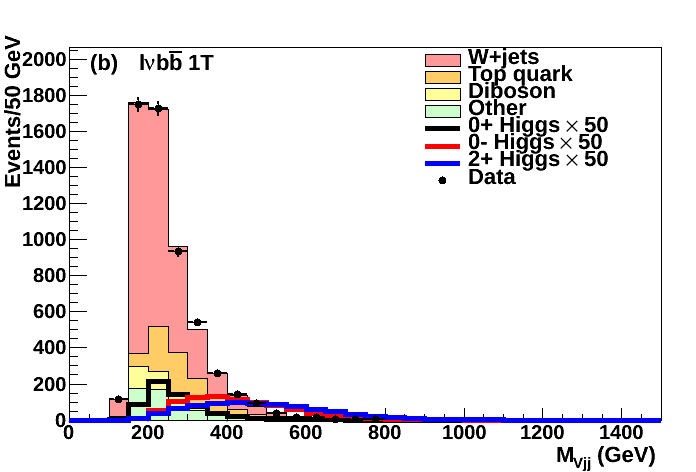}
 \caption{Reconstructed $Vb\bar b$ transverse invariant mass distribution
          for (a) the loose-loose and (b) the 1-tight $b$-jet categories of
          the $\ell\nu b{\bar{b}}$ search channel.}
 \end{centering}
 \end{figure*}

\begin{figure*}[htb]
 \begin{centering}
\includegraphics[width=0.45\linewidth]{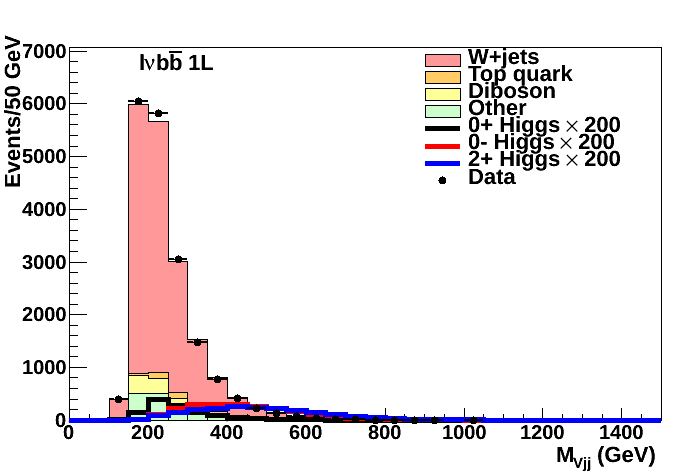}
 \caption{Reconstructed $Vb\bar b$ transverse invariant mass distribution
          for the 1-loose $b$-jet category of the $\ell\nu b{\bar{b}}$ search
          channel.}
 \end{centering}
 \end{figure*}

\begin{figure*}[htb]
 \begin{centering}
\includegraphics[width=0.45\linewidth]{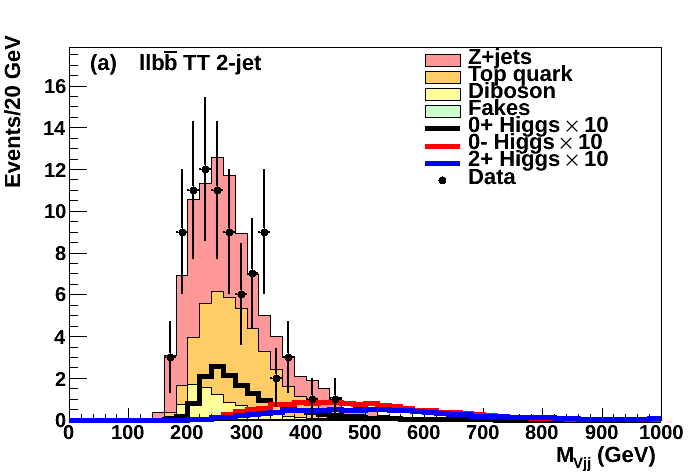}
\includegraphics[width=0.45\linewidth]{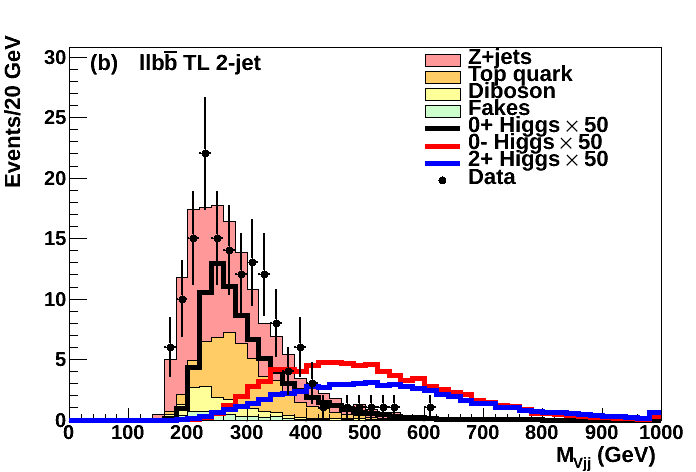}
 \caption{Reconstructed $Vb\bar b$ invariant mass distribution for (a)
          the tight-tight and (b) the tight-loose (right) $b$-jet categories
          in the 2-jet bin of the $\ell\ell b{\bar{b}}$ search channel.}
 \end{centering}
 \end{figure*}

\begin{figure*}[htb]
 \begin{centering}
\includegraphics[width=0.45\linewidth]{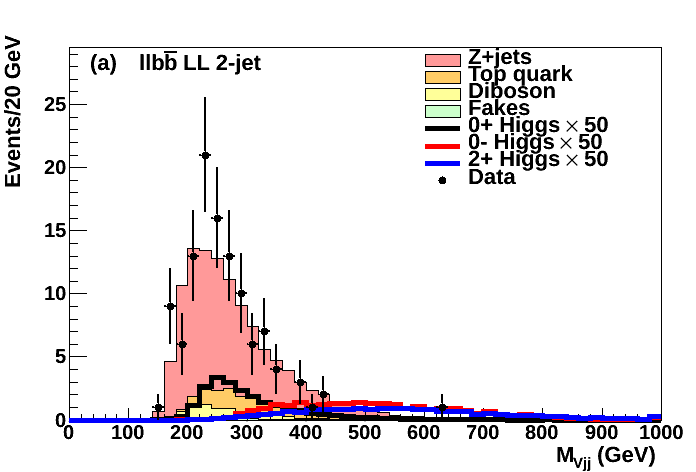}
\includegraphics[width=0.45\linewidth]{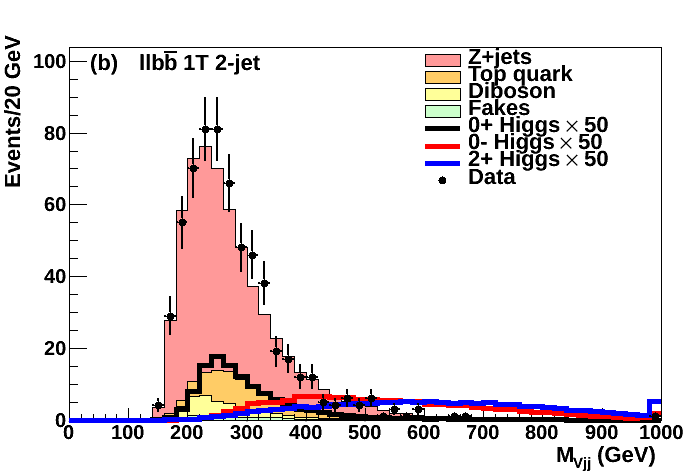}
 \caption{Reconstructed $Vb\bar b$ invariant mass distribution for (a)
          the loose-loose and (b) the 1-tight $b$-jet categories in the
          2-jet bin of the $\ell\ell b{\bar{b}}$ search channel.}
 \end{centering}
 \end{figure*}

\begin{figure*}[htb]
 \begin{centering}
\includegraphics[width=0.45\linewidth]{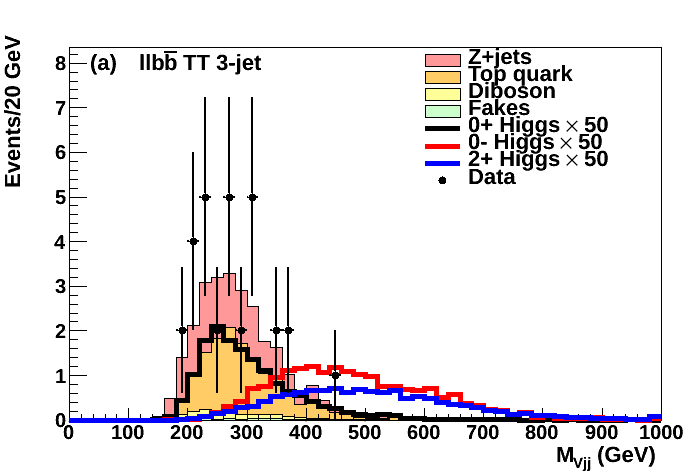}
\includegraphics[width=0.45\linewidth]{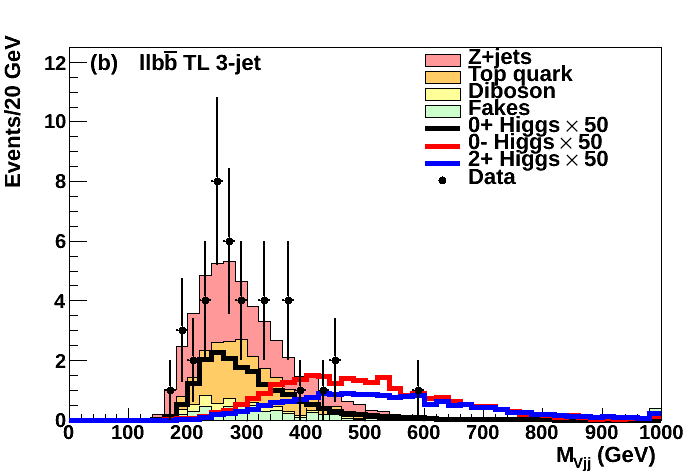}
 \caption{Reconstructed $Vb\bar b$ invariant mass distribution for (a)
          the tight-tight (left) and (b) the tight-loose $b$-jet categories
          in the 3-jet bin of the $\ell\ell b{\bar{b}}$ search channel.}
 \end{centering}
 \end{figure*}

\begin{figure*}[htb]
 \begin{centering}
\includegraphics[width=0.45\linewidth]{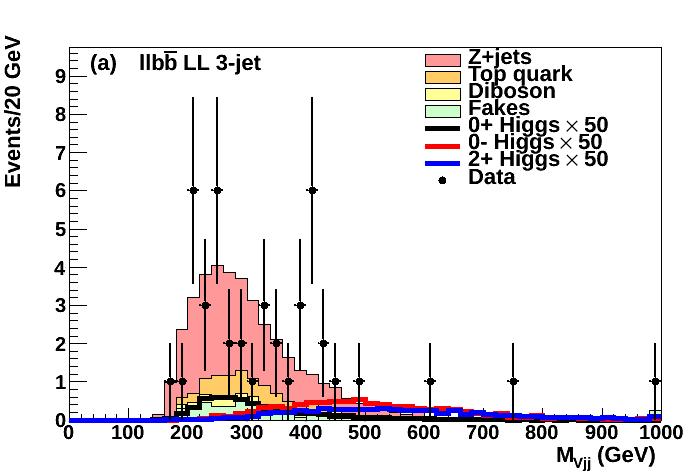}
\includegraphics[width=0.45\linewidth]{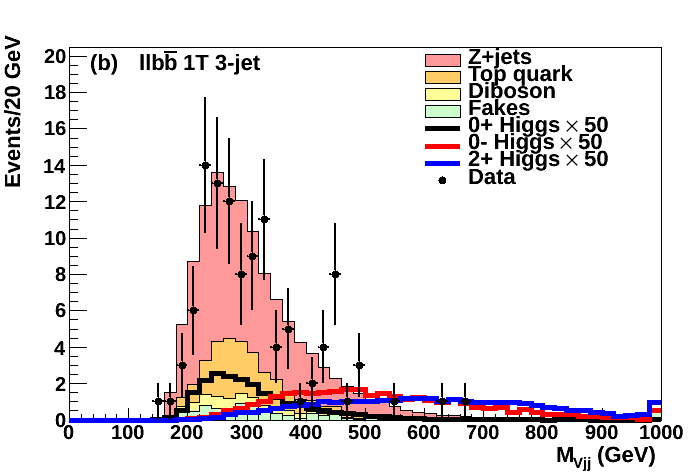}
 \caption{Reconstructed $Vb\bar b$ invariant mass distribution for (a)
          the loose-loose (left) and (b) the 1-tight $b$-jet categories
          in the 3-jet bin of the $\ell\ell b{\bar{b}}$ search channel.}
 \end{centering}
 \end{figure*}

\begin{figure*}[htb]
 \begin{centering}
\includegraphics[width=0.45\linewidth]{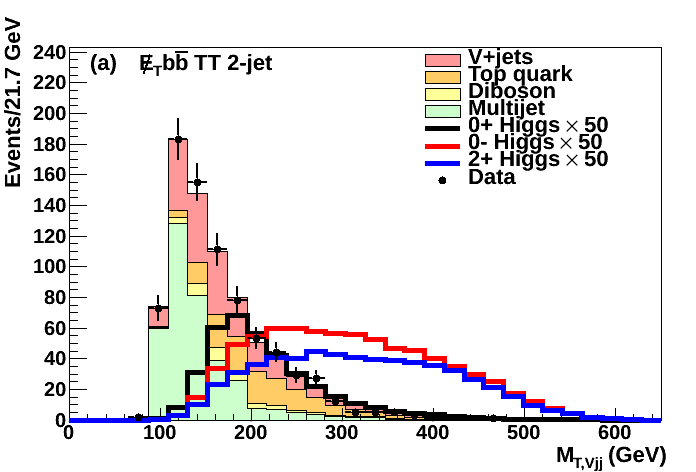}
\includegraphics[width=0.45\linewidth]{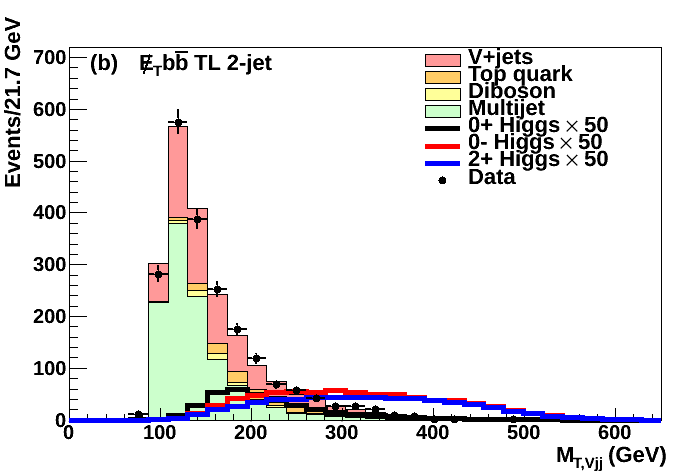}
 \caption{Reconstructed $Vb\bar b$ transverse invariant mass distribution
          for (a) the tight-tight and (b) the tight-loose $b$-jet categories
          in the 2-jet bin of the $\met b{\bar{b}}$ search channel.}
 \end{centering}
 \end{figure*}

\begin{figure*}[htb]
 \begin{centering}
\includegraphics[width=0.45\linewidth]{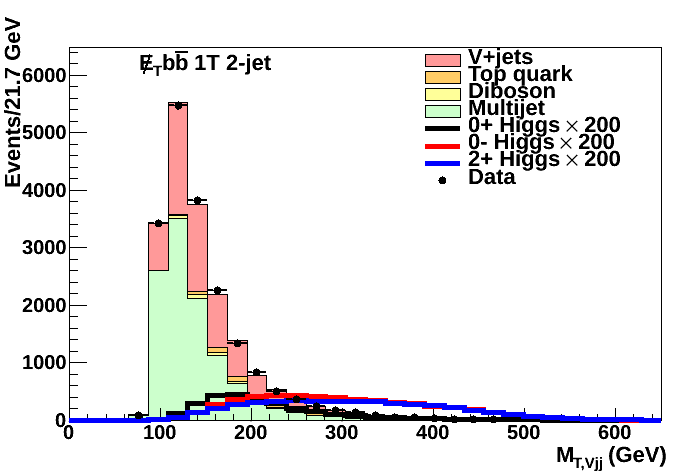}
 \caption{Reconstructed $Vb\bar b$ transverse invariant mass distribution
          for the 1-tight $b$-jet category in the 2-jet bin of the
          $\met b{\bar{b}}$ search channel.}
 \end{centering}
 \end{figure*}

\begin{figure*}[htb]
 \begin{centering}
\includegraphics[width=0.45\linewidth]{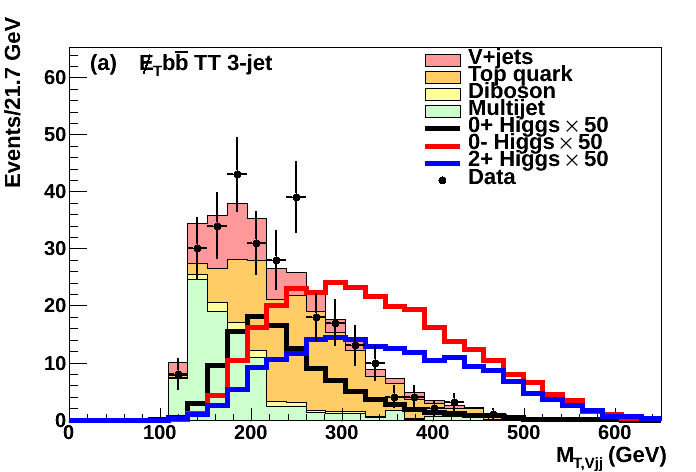}
\includegraphics[width=0.45\linewidth]{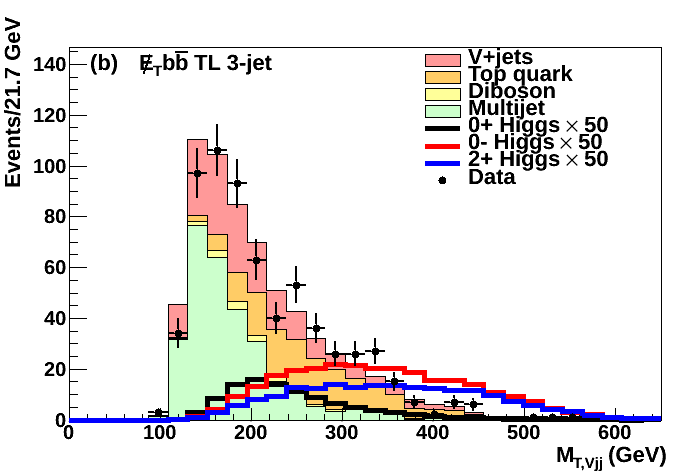}
 \caption{Reconstructed $Vb\bar b$ transverse invariant mass distribution
          for (a) the tight-tight and (b) the tight-loose $b$-jet categories
          in the 3-jet bin of the $\met b{\bar{b}}$ search channel.}
 \end{centering}
 \end{figure*}

\begin{figure*}[htb]
 \begin{centering}
\includegraphics[width=0.45\linewidth]{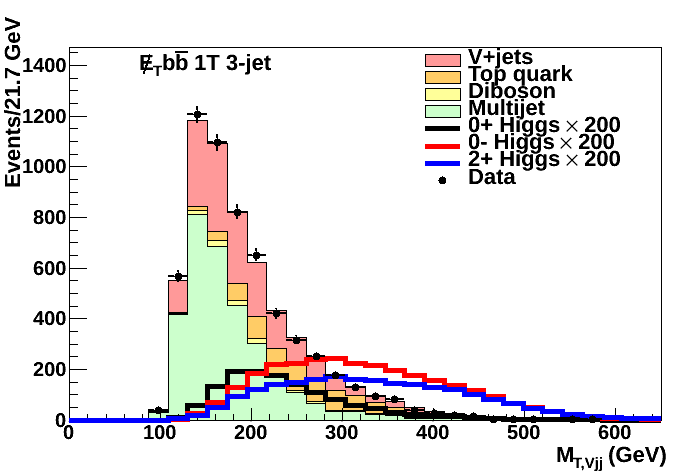}
 \caption{Reconstructed $Vb\bar b$ transverse invariant mass distribution
          for the 1-tight $b$-jet category in the 3-jet bin of the
          $\met b{\bar{b}}$ search channel.}
 \end{centering}
 \end{figure*}

\clearpage

\begin{figure*}[htb]
 \begin{centering}
\includegraphics[width=0.45\linewidth]{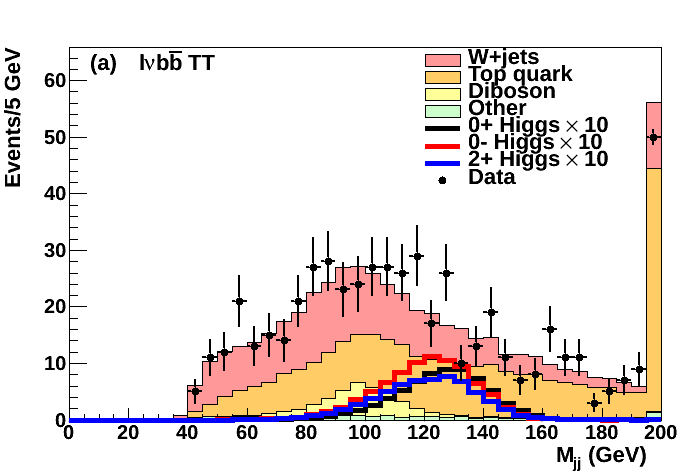}
 \includegraphics[width=0.45\linewidth]{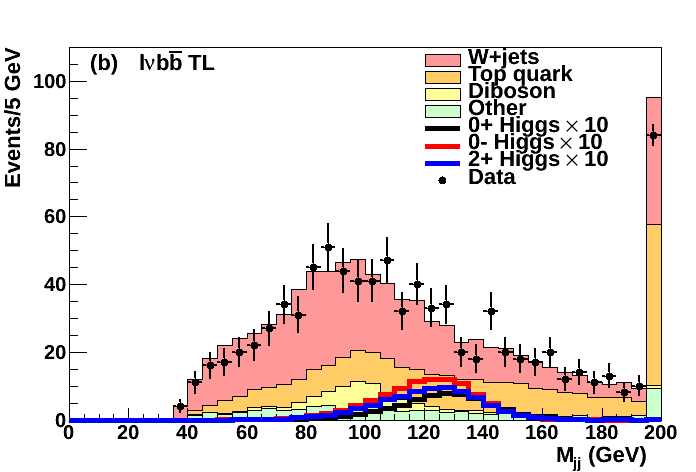}
 \caption{Reconstructed $b\bar b$ invariant mass distribution for (a)
          the tight-tight and (b) the tight-loose $b$-jet categories of the
          $\ell\nu b{\bar{b}}$ search channel.  Overflows are added to the
          content of the uppermost bin.}
 \end{centering}
 \end{figure*}

\begin{figure*}[htb]
 \begin{centering}
\includegraphics[width=0.45\linewidth]{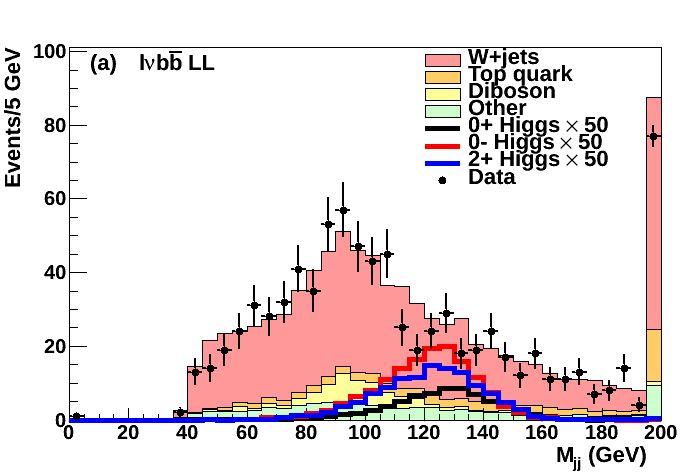}
 \includegraphics[width=0.45\linewidth]{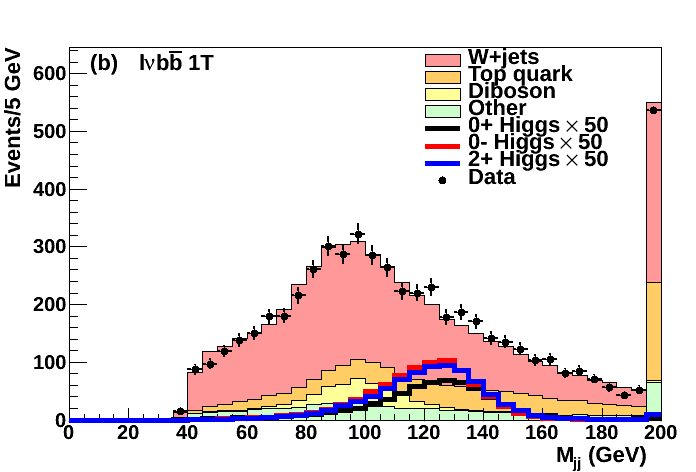}
 \caption{Reconstructed $b\bar b$ invariant mass distribution for (a)
          the loose-loose and (b) the 1-tight $b$-jet categories of the
          $\ell\nu b{\bar{b}}$ search channel.  Overflows are added to the
          content of the uppermost bin.}
 \end{centering}
 \end{figure*}

\begin{figure*}[htb]
 \begin{centering}
\includegraphics[width=0.45\linewidth]{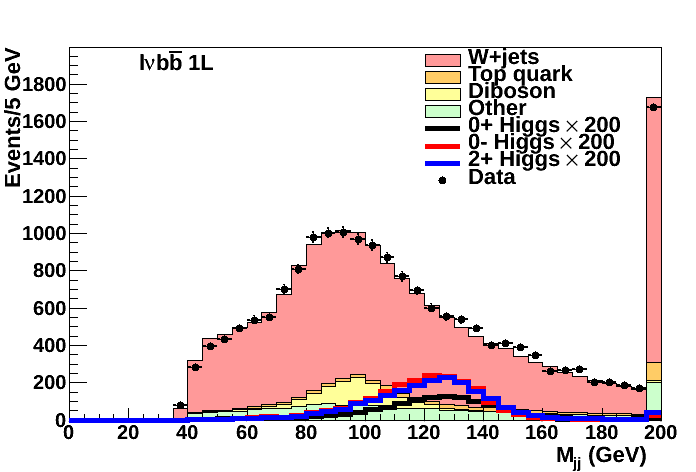}
 \caption{Reconstructed $b\bar b$ invariant mass distribution for the
          1-loose $b$-jet category of the $\ell\nu b{\bar{b}}$ search channel.
          Overflows are added to the content of the uppermost bin.}
 \end{centering}
 \end{figure*}

\begin{figure*}[htb]
 \begin{centering}
\includegraphics[width=0.45\linewidth]{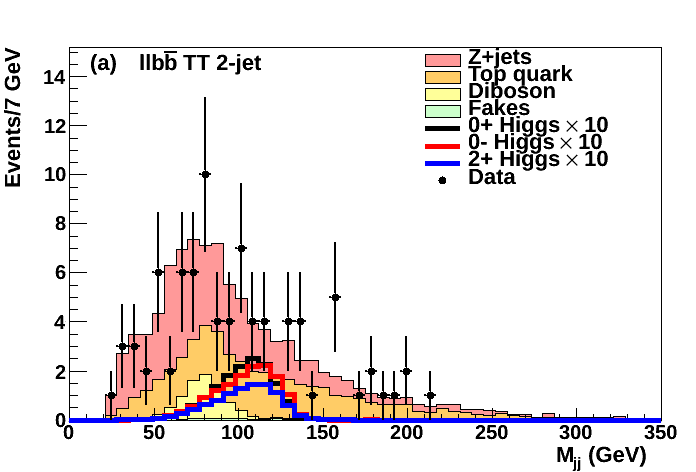}
\includegraphics[width=0.45\linewidth]{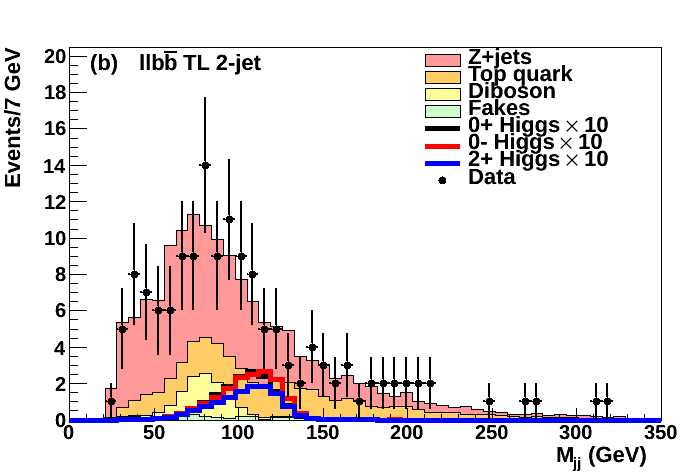}
 \caption{Reconstructed $b\bar b$ invariant mass distribution for (a)
          the tight-tight and (b) the tight-loose $b$-jet categories
          in the 2-jet bin of the $\ell\ell b{\bar{b}}$ search channel.}
 \end{centering}
 \end{figure*}

\begin{figure*}[htb]
 \begin{centering}
\includegraphics[width=0.45\linewidth]{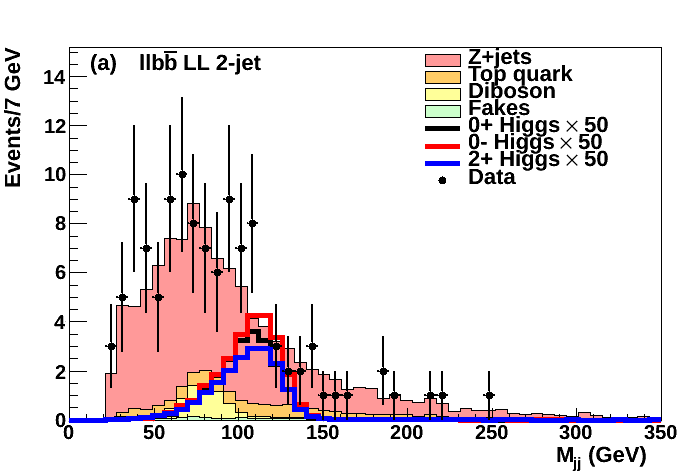}
\includegraphics[width=0.45\linewidth]{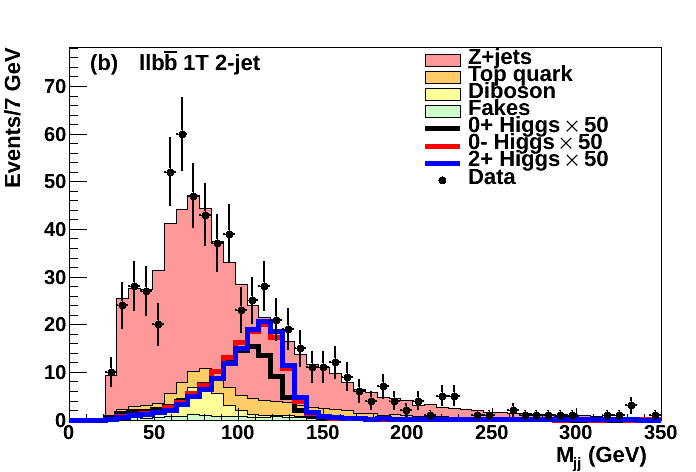}
 \caption{Reconstructed $b\bar b$ invariant mass distribution for (a)
          the loose-loose and (b) the 1-tight $b$-jet categories in the
          2-jet bin of the $\ell\ell b{\bar{b}}$ search channel.}
 \end{centering}
 \end{figure*}

\begin{figure*}[htb]
 \begin{centering}
\includegraphics[width=0.45\linewidth]{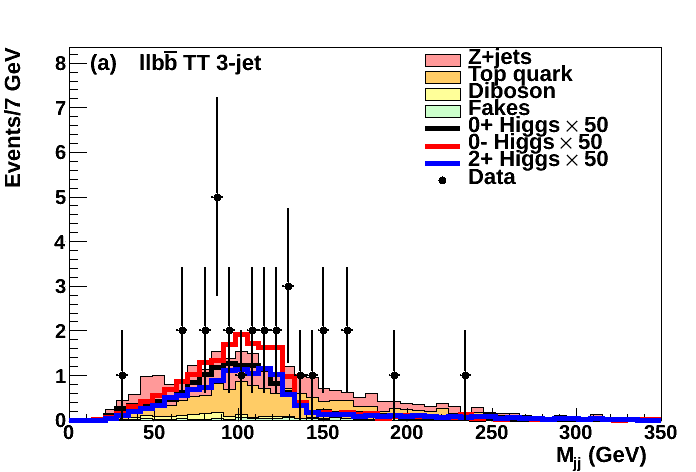}
\includegraphics[width=0.45\linewidth]{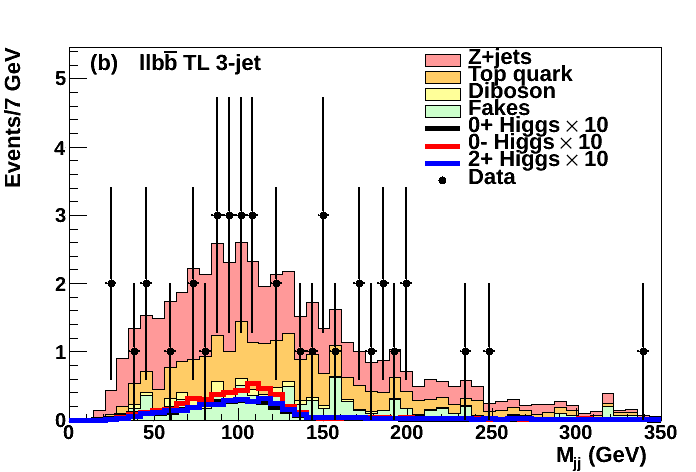}
 \caption{Reconstructed $b\bar b$ invariant mass distribution for (a)
          the tight-tight and (b) the tight-loose $b$-jet categories
          in the 3-jet bin of the $\ell\ell b{\bar{b}}$ search channel.}
 \end{centering}
 \end{figure*}

\begin{figure*}[htb]
 \begin{centering}
\includegraphics[width=0.45\linewidth]{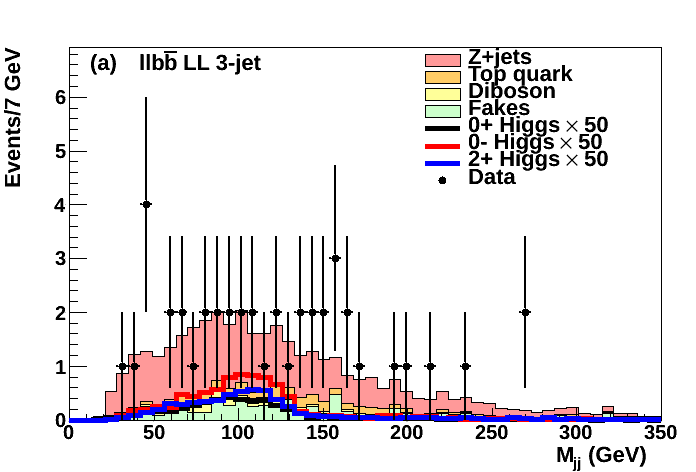}
\includegraphics[width=0.45\linewidth]{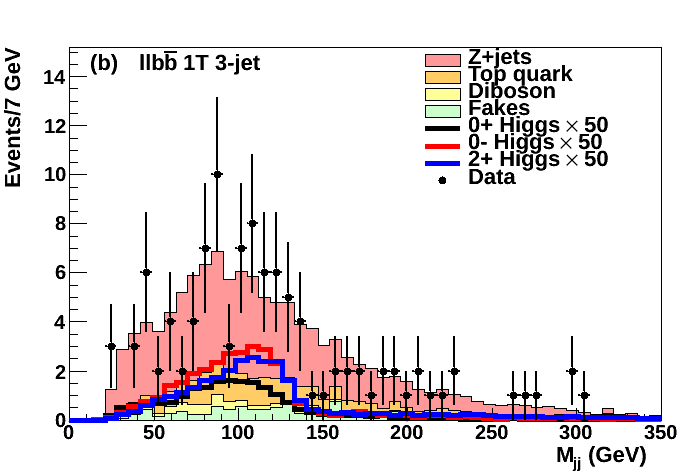}
 \caption{Reconstructed $b\bar b$ invariant mass distribution for (a)
          the loose-loose and (b) the 1-tight $b$-jet categories in the
          3-jet bin of the $\ell\ell b{\bar{b}}$ search channel.}
 \end{centering}
 \end{figure*}

\begin{figure*}[htb]
 \begin{centering}
\includegraphics[width=0.45\linewidth]{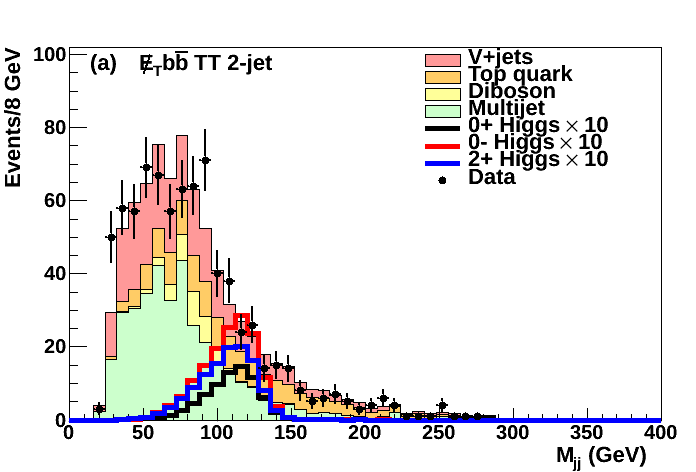}
\includegraphics[width=0.45\linewidth]{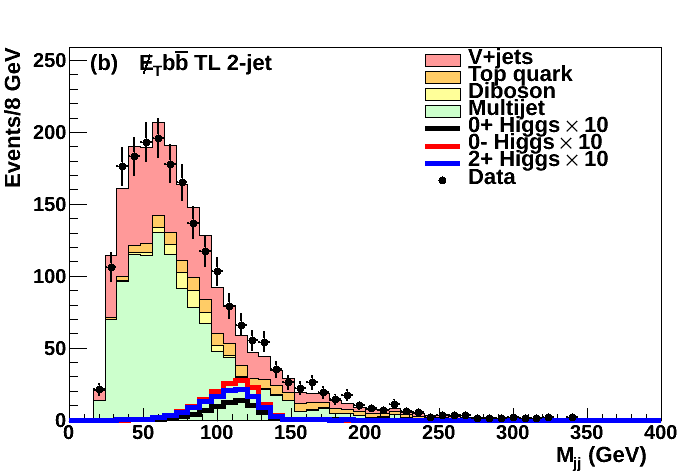}
 \caption{Reconstructed $b\bar b$ invariant mass distribution for (a)
          the tight-tight and (b) the tight-loose $b$-jet categories
          in the 2-jet bin of the $\met b{\bar{b}}$ search channel.}
 \end{centering}
 \end{figure*}

\begin{figure*}[htb]
 \begin{centering}
\includegraphics[width=0.45\linewidth]{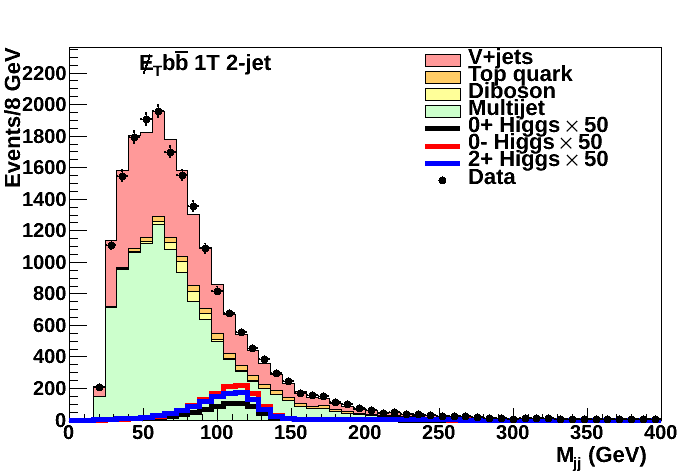}
 \caption{Reconstructed $b\bar b$ invariant mass distribution for the
          1-tight $b$-jet category in the 2-jet bin of the $\met b{\bar{b}}$
          search channel.}
 \end{centering}
 \end{figure*}

\begin{figure*}[htb]
 \begin{centering}
\includegraphics[width=0.45\linewidth]{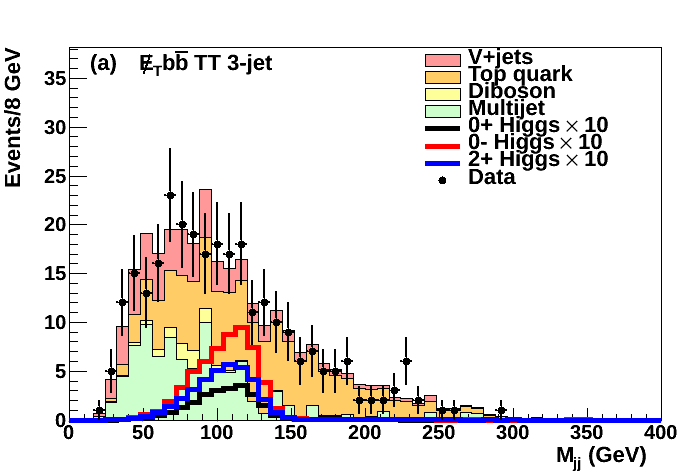}
\includegraphics[width=0.45\linewidth]{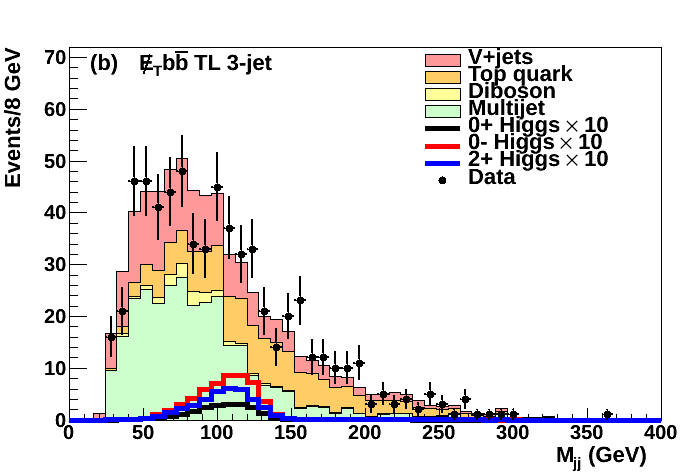}
 \caption{Reconstructed $b\bar b$ invariant mass distribution for (a)
          the tight-tight and (b) the tight-loose $b$-jet categories
          in the 3-jet bin of the $\met b{\bar{b}}$ search channel.}
 \end{centering}
 \end{figure*}

\begin{figure*}[htb]
 \begin{centering}
\includegraphics[width=0.45\linewidth]{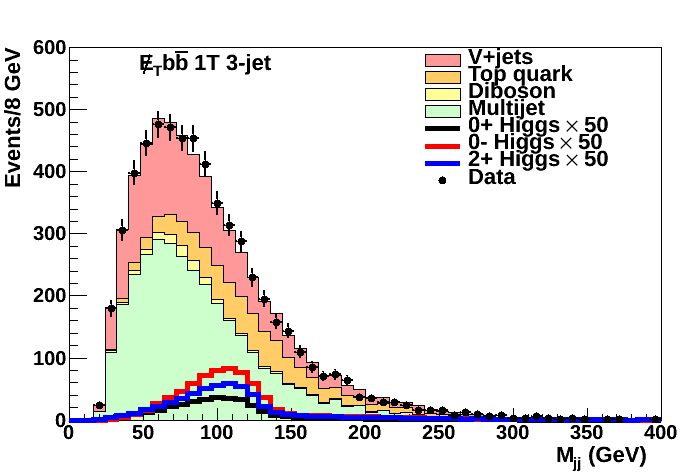}
 \caption{Reconstructed $b\bar b$ invariant mass distribution for the
          1-tight $b$-jet category in the 3-jet bin of the $\met b{\bar{b}}$
          search channel.}
 \end{centering}
 \end{figure*}


\begin{thebibliography}{99}

\bibitem{Aad:2012tfa} 
  G.~Aad {\it et al.}  (ATLAS Collaboration),
  Phys.\ Lett.\ B {\bf 716}, 1 (2012).

\bibitem{Chatrchyan:2012ufa} 
  S.~Chatrchyan {\it et al.}  (CMS Collaboration),
  Phys.\ Lett.\ B {\bf 716}, 30 (2012).

\bibitem{Aaltonen:2012qt} 
  T.~Aaltonen {\it et al.}  (CDF and D0 Collaborations),
  Phys.\ Rev.\ Lett.\  {\bf 109}, 071804 (2012).

\bibitem{Aaltonen:2013kxa} 
  T.~Aaltonen {\it et al.}  (CDF and D0 Collaborations),
  Phys.\ Rev.\ D {\bf 88}, 052014 (2013).

\bibitem{Aad:2013wqa} 
  G.~Aad {\it et al.}  (ATLAS Collaboration),
  Phys.\ Lett.\ B {\bf 726}, 88 (2013).

\bibitem{Chatrchyan:2013lba} 
  S.~Chatrchyan {\it et al.}  (CMS Collaboration),
  J.\ High Energy Phys.\ 06 (2013) 081.

\bibitem{pdghiggs}
{\it Status of Higgs Boson Physics}, in 
 K.~A.~Olive {\it et al.} (Particle Data Group),
  Chin. Phys. C {\bf 38}, 090001 (2014).

\bibitem{Aad:2013xqa} 
  G.~Aad {\it et al.}  (ATLAS Collaboration),
  Phys.\ Lett.\ B {\bf 726}, 120 (2013).

\bibitem{Chatrchyan:2012jja} 
  S.~Chatrchyan {\it et al.}  (CMS Collaboration),
  Phys.\ Rev.\ Lett.\  {\bf 110}, 081803 (2013).

\bibitem{Aad:2015} 
  G.~Aad {\it et al.}  (ATLAS Collaboration),
  arXiv:hep-ex/1501.04943 (2015).  Submitted to J.\ High Energy Phys.

\bibitem{Chatrchyan:2014vua} 
  S.~Chatrchyan {\it et al.}  (CMS Collaboration),
  Nature Phys.\  {\bf 10} (2014).

\bibitem{Abazov:2014} 
  V.~M.~Abazov {\it et al.}  (D0 Collaboration),
  Phys.\ Rev.\ Lett.\  {\bf 113}, 161802 (2014).

\bibitem{Ellis:2012xd} 
  J.~Ellis, D.~S.~Hwang, V.~Sanz, and T.~You,
  J.\ High Energy Phys.\ 11 (2012) 134.

\bibitem{Miller:2001bi} 
  D.~J.~Miller, S.~Y.~Choi, B.~Eberle, M.~M.~M\"uhlleitner, and P.~M.~Zerwas,
  Phys.\ Lett.\ B {\bf 505}, 149 (2001).

\bibitem{footnote1}
This is the approximate $s/b$ ratio in CDF's
$WH\rightarrow\ell\nu b{\bar{b}}$ search with two jets and two tight
$b$-tags~\cite{Aaltonen:2012ic}.

\bibitem{Aaltonen:2012ic} 
  T.~Aaltonen {\it et al.}  (CDF Collaboration),
  Phys.\ Rev.\ Lett.\  {\bf 109}, 111804 (2012).

\bibitem{Aaltonen:2012id} 
  T.~Aaltonen {\it et al.}  (CDF Collaboration),
  Phys.\ Rev.\ Lett.\  {\bf 109}, 111803 (2012).

\bibitem{footnote2}
The missing transverse energy, measuring the total transverse energy imbalance
in an event, is defined by
${\not\!\! E}_{T}=\vert\sum_{\rm towers}{E}_{T}\hat{n}_{T}\vert$, where
$\hat{n}_{T}$ is the unit vector normal to the beam and pointing to a given
calorimeter tower, and ${E}_{T}$ is the transverse energy measured in that
tower~\cite{coord}.

\bibitem{Aaltonen:2013js} 
  T.~Aaltonen {\it et al.}  (CDF Collaboration),
  Phys.\ Rev.\ D {\bf 87}, 052008 (2013).

\bibitem{secvtx}  D.~Acosta {\it et al.} (CDF Collaboration), Phys. Rev. D
  {\bf 71}, 052003 (2005).

\bibitem{cdfdetector} A. Abulencia, {\it et al.}, J.\ Phys.\ G: Nucl.\ Part.\
  Phys.\ {\bf 34}, 2457 (2007).

\bibitem{cdfsilicon}  T. Aaltonen {\it et al.}, Nucl. Instrum. Methods A
  {\bf 729}, 153 (2013).

\bibitem{coord} Positions and angles are expressed in a cylindrical coordinate
system, with the $z$~axis directed along the proton beam.  The azimuthal angle
$\phi$ around the beam axis is defined with respect to a horizontal line
pointing outwards from the center of the Tevatron, and radii are measured with
respect to the beam axis.  The polar angle $\theta$ is defined with respect to
the proton beam direction, and the pseudorapidity $\eta$ is defined to be
$\eta=-\ln\left[\tan(\theta/2)\right]$.  The transverse energy (as measured by
the calorimeters) and momentum (as measured by the tracking systems) of a
particle are defined as $E_{\rm T}=E\sin\theta$ and $p_{\rm T}=p\sin\theta$,
respectively.

\bibitem{cdfmuons} G. Ascoli {\it et al.}, Nucl. Instrum. Methods A {\bf 268},
  33 (1988).

\bibitem{CLC} D. Acosta {\it et al.}, Nucl. Instrum. Methods A {\bf 494}, 57
  (2002). 

\bibitem{XFT}  E.~J. Thomson {\it et al.}, IEEE Trans. on Nucl. Science.
  {\bf 49}, 1063 (2002).

\bibitem{cdfl3} G.~Gomez-Ceballos {\it et al.}, Nucl. Instrum. Methods A
  {\bf 518}, 522 (2004).

\bibitem[{pyt()}]{pythia}
\bibinfo{note}{T.~Sjostrand, S.~Mrenna, and P.~Skands, J.\ High Energy Phys.\
  05 (2006) 026. We use \textsc{pythia}~version 6.216 to generate the
  Higgs boson signals.}

\bibitem[{cte()}]{cteq}
\bibinfo{note}{H.~L.~Lai {\it et al.}, Eur. Phys. J. C~{\bf 12}, 375 (2000).}

\bibitem{madgraph}  
  J.~Alwall, M.~Herquet, F.~Maltoni, O.~Mattelaer, and T.~Stelzer,
  J.\ High Energy Phys.\ 06 (2011) 128; \\
J.~Alwall, R.~Frederix, S.~Frixione, V.~Hirschi, F.~Maltoni, O.~Mattelaer,
  H.-S.~Shao, T.~Stelzer, P.~Torrielli, and M.~Zaro,
  J.\ High Energy Phys.\ 07 (2014) 079.

\bibitem{djouadibaglio}
J.~Baglio and A.~Djouadi,
 J.\ High Energy Phys.\ 10 (2010) 064.

\bibitem{v2hv} The Fortran program can be found on Michael Spira's web page
{\tt http://people.web.psi.ch/$\sim$mspira/proglist.html}, using the formulae
presented in T. Han and S. Willenbrock, Phys. Lett. B {\bf 273}, 167 (1991).

\bibitem{vhnnloqcd} O.~Brein, A.~Djouadi, and R.~Harlander, Phys. Lett. B
{\bf 579}, 149 (2004).

\bibitem{vhewcorr} M.~L.~Ciccolini,  S.~Dittmaier, and M.~Kramer, Phys.
Rev. D {\bf 68}, 073003 (2003).

\bibitem{lhcxs}
S.~Dittmaier {\it et al.} (LHC Higgs Cross Section Working Group
Collaboration), arXiv:hep-ph/1101.0593.

\bibitem{lhcdifferential}
S.~Dittmaier {\it et al.} (LHC Higgs Cross Section Working Group
Collaboration), arXiv:hep-ph/1201.3084.

\bibitem[{hde()}]{hdecay}
\bibinfo{note}{A.~Djouadi, J.~Kalinowski, and M.~Spira, Comput.\ Phys.\
  Commun.\ {\bf 108}, 56 (1998).}

\bibitem[{pro()}]{prophecy4f}
\bibinfo{note}{A.~Bredenstein, A.~Denner, S.~Dittmaier, and M.~M.~Weber, Phys.
  Rev. D {\bf 74}, 013004 (2006); A. Bredenstein, A.~Denner, S.~Dittmaier, A.
  M\H{u}ck, and M.~M.~Weber, J.\ High Energy Phys.\ 02 (2007) 080.}

\bibitem{dblittlelhc}
J.~Baglio and A.~Djouadi,
 J.\ High Energy Phys.\ 03 (2011) 055.

\bibitem{denner11}
A.~Denner, S.~Heinemeyer, I.~Puljak, D.~Rebuzzi, and M.~Spira,
 Eur. Phys. J. C {\bf 71}, 1753 (2011).

\bibitem[{mcf()}]{mcfm}
\bibinfo{note}{J.~M.~Campbell and R.~K.~Ellis, Phys.\ Rev.\ D {\bf 60}, 113006
  (1999).}

\bibitem[{moc()}]{mochuwer}
\bibinfo{note}{S.~Moch and P.~Uwer, Nucl.\ Phys.\ Proc.\ Suppl.\ {\bf 183}, 75
  (2008).}

\bibitem[{MST()}]{MSTW}
\bibinfo{note}{A.~D.~Martin, W.~J.~Stirling, R.~S.~Thorne, and G.~Watt, Eur.\
  Phys.\ J.\ C {\bf 63}, 189 (2009).}

\bibitem[{kid()}]{kidonakis_st}
\bibinfo{note}{N.~Kidonakis, Phys.\ Rev.\ D {\bf 74}, 114012 (2006).}

\bibitem[{Man()}]{Mangano:2002ea}
\bibinfo{note}{M.~Mangano, M.~Moretti, F.~Piccinini, R.~Pittau, and A.~Polosa,
  J.\ High Energy Phys.\ 07 (2003) 001.}

\bibitem{Freeman:2012uf}
J.~Freeman, T.~Junk, M.~Kirby, Y.~Oksuzian, T.~J.~Phillips, F.~D.~Snider,
  M.~Trovato, J.~Vizan, and W.~M.~Yao,
  Nucl.\ Instrum.\ Meth.\ A {\bf 697}, 64 (2013)

\bibitem[{vht()}]{vhtheory}
  O.~Brein, R.~V.~Harlander, M.~Weisemann, and T.~Zirke, Eur. Phys. J. C
  {\bf 72}, 1868 (2012).

\bibitem[{Bha()}]{Bhatti:2005ai}
\bibinfo{note}{A.~Bhatti {\it et al.}, Nucl.\ Instrum.\ Methods\ A {\bf 566},
  375 (2006).}

\bibitem[{ine()}]{inelppbarxs}
\bibinfo{note}{S.~Klimenko, J.~Konigsberg, and T.~M.~Liss, FERMILAB-FN-0741
  (2003).}

\bibitem[{pdg()}]{pdgstats}
\bibinfo{note}{{\it Statistics}, in K. Nakamura {\it et al.}
(Particle Data Group), J. Phys. G {\bf 37}, 075021 (2010).}

\end{thebibliography}
\end{document}